\documentclass[fleqn,usenatbib]{mnras}

\usepackage{newtxtext,newtxmath}

\usepackage[T1]{fontenc}

\DeclareRobustCommand{\VAN}[3]{#2}
\let\VANthebibliography\thebibliography
\def\thebibliography{\DeclareRobustCommand{\VAN}[3]{##3}\VANthebibliography}

\usepackage{bm}
\usepackage{tikz}
\usetikzlibrary{patterns}
\usetikzlibrary{positioning}
\usetikzlibrary{decorations.text}
\usetikzlibrary{external}
\usepackage{pgfplots}
\usepgfplotslibrary{fillbetween}
\usepackage{subfig}

\usepackage{tikz}
\usetikzlibrary{positioning}
\usetikzlibrary{shadows}

\usepackage{graphicx}	
\usepackage{amsmath}	

\usepackage{siunitx}
\DeclareSIUnit\kpc{kpc}
\DeclareSIUnit\Mpc{Mpc}
\DeclareSIUnit\Gpc{Gpc}
\DeclareSIUnit\Gyr{Gyr}

\DeclareSIUnit{\hub}{\mathit{h}}
\DeclareSIUnit{\invhub}{\mathit{h}^{-1}}
\newcommand{\SIF}[2]{\SI[parse-numbers=false]{#1}{#2}}

\title[Dependence of baryonic feedback on cosmology]{The FLAMINGO project: the coupling between baryonic feedback and cosmology in light of the $S_8$ tension}

\author[Elbers et al.]{Willem Elbers,$^{1}$\thanks{willem.h.elbers@durham.ac.uk} {Carlos S. Frenk,$^{1}$ Adrian Jenkins,$^{1}$ Baojiu Li,$^{1}$ John C. Helly,$^{1}$ Roi Kugel,$^{2}$}\newauthor {Matthieu Schaller,$^{2,3}$ Joop Schaye,$^{2}$ Joey Braspenning,$^{2}$ Juliana Kwan,$^{4,5}$ Ian G. McCarthy,$^{4}$}\newauthor {Jaime Salcido,$^{4}$ Marcel P. van Daalen,$^{2}$ Bert Vandenbroucke,$^{2}$ Silvia Pascoli$^{6,7}$}\\~\\
$^{1}$Institute for Computational Cosmology, Department of Physics, Durham University, South Road, Durham, DH1 3LE, UK\\
{$^{2}$Leiden Observatory, Leiden University, PO Box 9513, 2300 RA Leiden, the Netherlands}\\
{$^{3}$Lorentz Institute for Theoretical Physics, Leiden University, PO box 9506, 2300 RA Leiden, the Netherlands}\\
{$^{4}$Astrophysics Research Institute, Liverpool John Moores University, Liverpool L3 5RF, UK}\\
{$^{5}$Department of Applied Mathematics and Theoretical Physics, University of Cambridge, Wilberforce Road, Cambridge, CB3 0WA, UK}\\
{$^{6}$Dipartimento di Fisica e Astronomia, Universit\`a di Bologna, via Irnerio 46, 40126 Bologna, Italy}\\
{$^{7}$INFN, Sezione di Bologna, viale Berti Pichat 6/2, 40127 Bologna, Italy}
}

\date{Last updated xx; in original form xx}

\pubyear{2023}

\pubyear{2023}

\definecolor{lightgrey}{RGB}{230, 230, 230}

\begin{document}
\label{firstpage}
\pagerange{\pageref{firstpage}--\pageref{lastpage}}
\maketitle

\begin{abstract}
Large-scale structure surveys have reported measurements of the density of matter, $\Omega_\mathrm{m}$, and the amplitude of clustering, $\sigma_8$, that are in tension with the values inferred from observations of the cosmic microwave background. While this may be a sign of new physics that slows the growth of structure at late times, strong astrophysical feedback processes could also be responsible. In this work, we argue that astrophysical processes are not independent of cosmology and that their coupling naturally leads to stronger baryonic feedback in cosmological models with suppressed structure formation or when combined with a mechanism that removes dark matter from halos. We illustrate this with two well-motivated extensions of the Standard Model known to suppress structure formation: massive neutrinos and decaying dark matter. Our results, based on the FLAMINGO suite of hydrodynamical simulations, show that the combined effect of baryonic and non-baryonic suppression mechanisms is greater than the sum of its parts, particularly for decaying dark matter. We also show that the dependence of baryonic feedback on cosmology can be modelled as a function of the ratio $f_\mathrm{b}/c^2_\mathrm{v}\sim f_\mathrm{b}/(\Omega_\mathrm{m}\sigma_8)^{1/4}$ of the universal baryon fraction, $f_\mathrm{b}$, to a velocity-based definition of halo concentration, $c^2_\mathrm{v}$, giving an accurate fitting formula for the baryonic suppression of the matter power spectrum. Although the combination of baryonic and non-baryonic suppression mechanisms can resolve the tension, the models with neutrinos and decaying dark matter are challenged by constraints on the expansion history.
\end{abstract}




\section{Introduction}

The success of the $\Lambda$CDM model indicates that signatures of new physics are likely to manifest either as small modifications to the $\Lambda$CDM prediction or in the relatively unexplored high-redshift r\'egime. As such, the model is under intense scrutiny at both the high-precision and high-redshift frontiers. Among a number of tensions and puzzling anomalies \citep[e.g.][]{abdalla22,peebles22} is a long-standing discrepancy between measurements of the matter density and amplitude of fluctuations on $\SIF{8\invhub}{\Mpc}$ scales\footnote{Described by the parameter combination $S_8=(\Omega_\text{m}/0.3)^{1/2}\,\sigma_8$, where $\Omega_\text{m}$ is the present density of matter and $\sigma_8$ the standard deviation of the present linear matter field averaged in spheres of radius $\SIF{8\invhub}{\Mpc}$.} obtained from large-scale structure probes, such as galaxy clustering, weak lensing, and the thermal Sunyaev-Zeldovich (tSZ) effect, and the values extrapolated from measurements of the cosmic microwave background (CMB) \citep[e.g.][]{planck18,nunes21,asgari22,amon22,secco22,abbott23,mccarthy23}. Further motivating the work at the high-precision frontier is the possibility to measure the sum of neutrino masses. The imprint of massive neutrinos could be detected by galaxy surveys such as DESI, Euclid, and LSST, even for the minimum value allowed by oscillation data, but this requires percent-level accuracy in large-scale structure measurements and predictions \citep[e.g.][]{brinckmann19,chudaykin19}.

Complicating these efforts is the fact that astrophysical processes, such as feedback from supernovae (SN) and active galactic nuclei (AGN), change the distribution of matter even on relatively large scales \citep[e.g.][]{vandaalen11,chisari19,schneider19,debackere20}. By heating and ejecting gas into the intergalactic medium, AGN feedback can suppress the power spectrum of matter fluctuations by $\mathcal{O}\big(10\%\big)$ on non-linear scales, $\SIF{1\hub}{\per\Mpc}<k<\SIF{10\hub}{\per\Mpc}$. On smaller scales, the power spectrum may be boosted by star formation and gas cooling, both processes allowing matter to contract (e.g. \citealt{debackere20}; but see e.g. \citealt{forouhar22}). A crucial question for the interpretation of large-scale structure observations concerns the coupling between baryonic physics and cosmology. If the choice of cosmological model determined only the distribution of dark matter halos, while the galaxies formed inside those halos were identical, then the effects of baryons and cosmology might be modelled independently. On the other hand, a non-trivial coupling between galaxy formation and cosmological processes should give rise to `non-factorizable corrections' to clustering statistics, introduced formally below. Note that there is a difference between the factorizability of two processes, meaning that their effects on some observable can be treated independently, and their degeneracy, meaning that their effects are indistinguishable.

A number of previous studies \citep[e.g.][]{vandaalen11,vandaalen20,mummery17,pfeifer20,schneider20,stafford20,arico21,parimbelli21,broxterman23,salcido23,upadhye23} have shown that the effects of cosmology and baryonic physics are indeed factorizable to a first approximation, with residual effects of up to several percent for small variations in cosmology. There are several reasons to subject this topic to further systematic scrutiny. First of all, by modelling more precisely the non-factorizable corrections that arise from variations in cosmology in the presence of baryonic physics, we can improve existing prescriptions for baryonic feedback \citep[e.g.][]{mead15,mead21} and match the precision of Stage-IV galaxy surveys such as Euclid and LSST. Second, new physics introduced to account for tensions in cosmological datasets may change more significantly the strength of baryonic effects than simply varying the standard cosmological parameters. For example, we will demonstrate a strong dependence of baryonic effects on the dark matter lifetime. Understanding the coupling between cosmology and baryonic feedback will shed light on the conditions under which their interaction becomes important and guide model builders towards novel solutions of cosmic tensions. Third, since baryonic feedback can itself be constrained in multiple independent ways, such as through weak lensing and X-ray or SZ measurements of cluster gas fractions \citep[e.g.][]{semboloni11,harnoisderaps15,mccarthy17,schneider20,schneider22,amon22,arico23,chen23,grandis23,kugel23,to24}, its dependence on cosmology might in principle be used as a cosmological probe if the degeneracy with astrophysical parameters and modelling uncertainty could be broken.

One could imagine different mechanisms through which baryonic processes, such as star formation and the growth of supermassive black holes, and hence baryonic feedback, could depend on cosmology. Although dark matter halos share a universal density profile \citep{navarro96b,navarro96}, their concentrations depend on cosmology \citep{eke01,knollmann08,prada11,kwan13,correa15,ragagnin21}. Cosmological model variations that slow the rate of structure formation (such as decreasing the matter density, $\Omega_\text{m}$, or amplitude of clustering, $\sigma_8$) lead to less concentrated halos, lowering the gravitational binding energy and altering the balance between outflows and black hole accretion \citep{booth10,bower17,chen20}. Another potential channel is the formation history of dark matter halos. If halos assemble their mass more slowly, galaxy formation and the rapid growth of supermassive black holes may be delayed \citep{matthee17,davies22}. A third possibility is that a change in the large-scale distribution of matter affects the halo environment, which could affect halo properties indirectly through assembly bias \citep{avilareese05,wechsler06,gao07} or affect feedback by changing the density of the halo outskirts. Finally and perhaps most crucially, variations in the baryon fraction, through shifts in $\Omega_{\text{b}}$, alter the amount of gas that is available for star formation and black hole accretion.

These effects must be considered when extensions of the $\Lambda$CDM model are introduced. Consider massive neutrinos as an example. A change in the sum of neutrino masses, $\sum m_\nu$, could plausibly affect feedback through any of the four channels mentioned above. Neutrinos cluster less effectively on scales smaller than their free streaming length \citep[for a review, see][]{lesgourgues06}, which results in less concentrated halos, delayed structure formation, and smoother halo environments. Moreover, neutrinos also affect the baryon density, $\Omega_\text{b}/\Omega_\text{c}$, relative to the cold dark matter density, $\Omega_{\text{c}}$, given that a change in neutrino mass at fixed matter density, $\Omega_\text{m}=\Omega_\text{b}+\Omega_\text{c}+\Omega_\nu$, will alter the amount of gas that is available for a halo of a given dark matter mass. Crucially, the universal baryon fraction should therefore be defined relative to the mass that clusters efficiently: $f_\mathrm{b}\equiv\Omega_\text{b}/(\Omega_\text{c} + \Omega_\text{b})$. These channels are not necessarily mutually exclusive. To find out which, if any, play a role in regulating baryonic feedback, we will use the new FLAMINGO suite of hydrodynamical simulations \citep{kugel23,schaye23}, which includes several feedback and cosmology variations. We will use halo properties (such as the concentration and formation epoch) as proxies for the ways in which feedback could depend on cosmology. We will then formulate a model to predict the non-factorizable corrections to the matter power spectrum in response to a shift in cosmology.

The organization of the paper is as follows. In Section \ref{sec:sims}, we discuss the FLAMINGO model and describe the simulations analysed in this paper. In Section \ref{sec:results}, we will first demonstrate the existence of non-factorizable corrections for models that are close to a Planck-based $\Lambda$CDM model. This is followed by an analysis of possible mediating mechanisms. Using insights from that analysis, we then develop an analytical model for the dependence of baryonic feedback on cosmology and test it against the simulations. Those readers primarily interested in the $S_8$ tension may on first reading skip to Section~\ref{sec:application}, where we discuss the cosmological implications of these results in light of this tension and extensions of the $\Lambda$CDM model. In Section~\ref{section:approximations}, we compare our results with commonly-used halo model approaches. Finally, Section \ref{sec:discussion} provides the conclusion.

{
\setlength{\tabcolsep}{5.5pt}
\begin{table*}
	\centering
	\caption{An overview of the large FLAMINGO simulations used in this paper. The number of baryon particles, $N_\text{b}$, is equal to the number of cold dark matter particles, $N_\text{c}$, for the simulations that have them. The number of neutrino particles is always $N_\nu=N_\text{c}/1.8^3$. For each simulation, there exists a gravity-only (DMO) counterpart with combined CDM and baryon particle mass $m_\text{cb} = m_\text{c} + m_\mathrm{g}$. The columns correspond to the side length, $L$, the number and (initial) mass of cold dark matter particles, $N_\text{c}$ and $m_\text{c}$, the initial mass of gas particles, $m_\mathrm{g}$, and the cosmological parameters. The final column shows the dark matter decay rate, $\Gamma$, in units of $\SI{100}{\km/\s/\Mpc}=H_0/h$ for the simulations with decaying cold dark matter (DCDM). For the DCDM models, the $\Omega_\text{c}$ column lists the sum of the present-day densities of decaying cold dark matter and dark radiation. The simulations with the suffix $f_\text{gas}\!\pm\!n\sigma$ are identical to L1\_m9, but have subgrid physics parameters calibrated to cluster gas fractions that are $n\sigma$ higher or lower than the observed data. All simulations assume a flat $(\Omega_\mathrm{k}=0)$ Universe with massive neutrinos and with an amount of radiation corresponding to $T_\text{CMB}=\SI{2.7255}{\K}$ and $N_\text{eff}=3.044$ effective relativistic neutrino species at high redshift.}
	\label{tab:sims}
	\begin{tabular}{lcccccccccccccccc}
		\hline
        Identifier & $L/\si{\Gpc}$ & $N_\text{c}$ & $m_\text{c}/\si{M_\odot}$ & $m_\mathrm{g}/\si{M_\odot}$ & $h$ & $\Omega_\text{m}$ & $\Omega_\text{c}$ & $\Omega_\text{b}$ & $\sum m_\nu$ & $\sigma_8$ & $10^9 A_\text{s}$ & $n_\text{s}$ & $\Gamma h/H_0$ \\
		\hline
		{\footnotesize L2p8\_m9}          & $2.8$ & $5040^3$ & $5.65\times10^9$ & $1.07\times10^9$ & $0.681$ & $0.306$ & $0.256$ & $0.0486$ & $\SI{0.06}{\eV}$ & $0.807$ & $2.099$ & $0.967$ & -- \\
		{\footnotesize L1\_m9}          & $1.0$ & $1800^3$ & $5.65\times10^9$ & $1.07\times10^9$ & $0.681$ & $0.306$ & $0.256$ & $0.0486$ & $\SI{0.06}{\eV}$ & $0.807$ & $2.099$ & $0.967$ & -- \\
        {\footnotesize L1\_m9\_$f_\text{gas}\!-\!8\sigma$}          & $1.0$ & $1800^3$ & $5.65\times10^9$ & $1.07\times10^9$ & $0.681$ & $0.306$ & $0.256$ & $0.0486$ & $\SI{0.06}{\eV}$ & $0.807$ & $2.099$ & $0.967$ & -- \\
        {\footnotesize L1\_m9\_$f_\text{gas}\!-\!4\sigma$}          & $1.0$ & $1800^3$ & $5.65\times10^9$ & $1.07\times10^9$ & $0.681$ & $0.306$ & $0.256$ & $0.0486$ & $\SI{0.06}{\eV}$ & $0.807$ & $2.099$ & $0.967$ & -- \\
        {\footnotesize L1\_m9\_$f_\text{gas}\!-\!2\sigma$}          & $1.0$ & $1800^3$ & $5.65\times10^9$ & $1.07\times10^9$ & $0.681$ & $0.306$ & $0.256$ & $0.0486$ & $\SI{0.06}{\eV}$ & $0.807$ & $2.099$ & $0.967$ & -- \\
        {\footnotesize L1\_m9\_$f_\text{gas}\!+2\sigma$}          & $1.0$ & $1800^3$ & $5.65\times10^9$ & $1.07\times10^9$ & $0.681$ & $0.306$ & $0.256$ & $0.0486$ & $\SI{0.06}{\eV}$ & $0.807$ & $2.099$ & $0.967$ & -- \\
		{\footnotesize Planck}          & $1.0$ & $1800^3$ & $5.72\times10^9$ & $1.07\times10^9$ & $0.673$ & $0.316$ & $0.265$ & $0.0494$ & $\SI{0.06}{\eV}$ & $0.812$ & $2.101$ & $0.966$ & -- \\
		{\footnotesize Planck$\nu$0.24Var} & $1.0$ & $1800^3$ & $5.67\times10^9$ & $1.06\times10^9$ & $0.662$ & $0.328$ & $0.271$ & $0.0510$ & $\SI{0.24}{\eV}$ & $0.772$ & $2.109$ & $0.968$ & -- \\
		{\footnotesize Planck$\nu$0.24Fix} & $1.0$ & $1800^3$ & $5.62\times10^9$ & $1.07\times10^9$ & $0.673$ & $0.316$ & $0.261$ & $0.0494$ & $\SI{0.24}{\eV}$ & $0.769$ & $2.101$ & $0.966$ & -- \\
        {\footnotesize Planck$\nu$0.48Fix} & $1.0$ & $1800^3$ & $5.62\times10^9$ & $1.07\times10^9$ & $0.673$ & $0.316$ & $0.256$ & $0.0494$ & $\SI{0.48}{\eV}$ & $0.709$ & $2.101$ & $0.966$ & -- \\
        {\footnotesize PlanckDCDM12}          & $1.0$ & $1800^3$ & $5.71\times10^9$ & $1.07\times10^9$ & $0.673$ & $0.274$ & $0.246$ & $0.0494$ & $\SI{0.06}{\eV}$ & $0.794$ & $2.101$ & $0.966$ & $$ $0.12$ \\
        {\footnotesize PlanckDCDM24}          & $1.0$ & $1800^3$ & $5.70\times10^9$ & $1.07\times10^9$ & $0.673$ & $0.239$ & $0.229$ & $0.0494$ & $\SI{0.06}{\eV}$ & $0.777$ & $2.101$ & $0.966$ & $$ $0.24$ \\
		{\footnotesize LS8}             & $1.0$ & $1800^3$ & $5.65\times10^9$ & $1.07\times10^9$ & $0.682$ & $0.305$ & $0.256$ & $0.0473$ & $\SI{0.06}{\eV}$ & $0.760$ & $1.836$ & $0.965$ & -- \\
    	\hline
	\end{tabular}
\end{table*}
}

\section{Simulations}\label{sec:sims}

Our analysis is based on the FLAMINGO suite of cosmological hydrodynamical simulations \citep{schaye23}. The FLAMINGO simulations use an updated version of the subgrid models adopted by the earlier OWLS \citep{schaye10}, Cosmo-OWLS \citep{lebrun14}, and BAHAMAS \citep{mccarthy17} projects. In a departure from its predecessors, the subgrid physics parameters were systematically calibrated by training emulators to predict key astrophysical quantities (the galaxy stellar mass function at $z=0$ and cluster gas fractions at low $z$) and comparing directly with observations \citep{kugel23}. The simulations also used higher-order multi-fluid initial conditions and a novel treatment of massive neutrinos \citep{hahn21,elbers21,elbers22}, both aimed at improving the accuracy of its large-scale structure predictions. Combined with the unprecedented volume of the simulations, these improvements make FLAMINGO ideal for precision cosmology applications.

The largest simulation contains $N_\text{c}=N_\text{b}=5040^3$ dark matter and baryon particles and $N_\nu=2800^3$ massive neutrino particles in a periodic $(\SI{2.8}{\Gpc})^3$ volume. This simulation assumes a fiducial $\Lambda$CDM cosmology with minimal neutrino masses, $\sum m_\nu=\SI{0.06}{\eV}$, and parameters based on the Dark Energy Survey Year 3 analysis of $3\times2$pt clustering, BAO, RSD, SNe Ia, and Planck CMB data \citep{abbott22}. In addition, the FLAMINGO suite contains many simulations with the same mass resolution in a $(\SI{1}{\Gpc})^3$ volume. These simulations span a range of subgrid physics and cosmological parameter variations. Among these are simulations with the fiducial cosmology, but with subgrid parameters calibrated to cluster gas fractions that are $n\sigma$ above or below the observations. For each hydrodynamical simulation, there is a gravity-only counterpart that treats dark matter and baryons as a single cold fluid, but still includes the effects of massive neutrinos (also called ``dark matter only'' or DMO).

The hydrodynamical simulations include improved prescriptions for gas cooling \citep{ploeckinger20}, star formation \citep{schaye08}, stellar feedback \citep{chaikin22}, and chemical enrichment \citep{wiersema09}. Supermassive black holes are modelled following the approaches of \citet{springel05}, \citet{dimatteo08}, \citet{booth09}, and \citet{bahe22}. For some runs, AGN feedback was implemented as jets instead of thermally driven winds \citep{husko22}. The simulations were run with the \textsc{swift} cosmological hydrodynamics code \citep{schaller18,schaller23}, using the SPHENIX flavour of SPH \citep{borrow22}, on the \textsc{cosma}-8 facility at Durham.

The initial conditions were generated with third-order Lagrangian perturbation theory (3LPT) at $z=31$ with separate transfer functions for dark matter, baryons, and neutrinos \citep{hahn21,elbers22}, using a modified version of \textsc{monofonIC}\footnote{\url{https://github.com/wullm/monofonic}} \citep{hahn20,michaux21} and \textsc{fastdf} \citep{elbers22b}, with transfer functions computed with \textsc{class} \citep{lesgourgues11}. Halo catalogues were produced with \textsc{velociraptor} \citep{elahi19} and post-processed with the spherical overdensity analysis tool \textsc{soap}\footnote{\url{https://github.com/SWIFTSIM/SOAP}}.

\subsection{Model extensions}

To explore the varying impact of cosmology-dependent feedback for models beyond $\Lambda$CDM, we consider two extensions of the base model. Although all simulations have neutrinos, we will consider runs that vary the massive neutrino content. In addition, we will consider models in which dark matter is unstable and decays invisibly to dark radiation \citep[e.g.][]{cen01,wang10,audren14,aoyama14,berezhiani15,enqvist15,poulin16,hubert21,tanimura23}. Both extensions retard the growth of structure compared to $\Lambda$CDM with minimal neutrino masses, which we will show also enhances the strength of baryonic feedback.

Massive neutrinos were included in the simulations using the $\delta f$ method \citep{elbers21}, which minimizes shot noise without neglecting the nonlinear evolution of the phase-space distribution. The neutrino particles start from relativistic initial conditions and use a special relativistic velocity correction to produce accurate clustering on large scales. For models with $\sum m_\nu=\SI{0.06}{\eV}$, one massive species and two massless species were assumed, with the latter contributing only at the background level. For larger neutrino masses, a degenerate mass spectrum was assumed. In \citet{elbers21,elbers22}, it is shown that the effects of neutrinos on the power spectrum can be modelled with $0.1\%$-level accuracy, enabling the detailed analysis in this paper (see also \citealt{adamek23}).

In the decaying cold dark matter model (DCDM), cold dark matter is an unstable particle with mean lifetime $\tau$ that decays into a new relativistic particle, referred to as dark radiation. In general, only a fraction $f$ of the dark matter might be unstable \citep[e.g.][]{berezhiani15,hubert21}, but we will focus on the $f=1$ case for simplicity. The background densities of decaying cold dark matter and dark radiation then evolve as
\begin{align}
\dot{\rho}_\text{dcdm} &= -3\mathcal{H}\rho_\text{dcdm} - a\Gamma_\text{dcdm}\rho_\text{dcdm},\\
\dot{\rho}_\text{dr} &= -4\mathcal{H}\rho_\text{dr} + a\Gamma_\text{dcdm}\rho_\text{dcdm},
\end{align}
\noindent
where $\Gamma_\text{dcdm}=\tau^{-1}$ is the decay rate, dots denote conformal time derivatives, $a$ is the scale factor, and $\mathcal{H}=\dot{a}/a$. These equations were implemented at the background level in the $N$-body code \textsc{swift}, modifying the equations of motion in the expanding reference frame. At the perturbation level, dark matter decay was implemented by adjusting the particle masses in proportion to the fraction of dark matter that has decayed, following \citet{enqvist15} and \citet{hubert21}. For the gravity-only (DMO) simulations, the evolving particle masses take into account the fact that $\Omega_\text{b}$ is constant. We also modified the initial conditions code \textsc{monofonIC} \citep{hahn20} and backscaling code \textsc{zwindstroom} \citep{elbers22c} to account for dark matter decay and used the publicly available implementation of decaying dark matter in \textsc{class}\footnote{\url{https://github.com/lesgourg/class_public}} \citep{audren14} for the transfer functions.

We ran two large hydrodynamical simulations with the DCDM model at the fiducial resolution in a $(\SI{1}{\Gpc})^3$ volume, using cosmological parameters based on the existing Planck simulation \citep{schaye23}, but adjusting $\Omega_\Lambda$ and the initial value of $\Omega_\text{dcdm}$ to keep the primordial dark matter to baryon density ratio and the curvature, $\Omega_\text{k} = 0$, fixed at the Planck values. Our choice for the decay rate, $\Gamma$, is based on an analysis of the final data release of Planck CMB temperature and polarization data by \citet{tristram23}. Using the \textsc{cobaya} sampler \citep{torrado21}, we obtain a $95\%$ upper bound of $\Gamma<\SI{12.8}{\km/\s/\Mpc}$. This is to be compared with a recent analysis by \citet{tanimura23} of CMB, tSZ, BAO, and SNIa data, which gave a $95\%$ upper bound of $\Gamma<\SI{26}{\km/\s/\Mpc}$ and a preferred value of $\Gamma=\SI{7.1}{\km/\s/\Mpc}$. On the other hand, tighter constraints of $\Gamma\lesssim\SI{6}{\km/\s/\Mpc}$ have been obtained from other data combinations \citep{audren14,aubourg15,enqvist20}.

Based on these results, we choose for our simulations decay rates of $\Gamma=\SI{12}{\km/\s/\Mpc}$ ($\tau=\SI{81}{\Gyr}$) and $\Gamma=\SI{24}{\km/\s/\Mpc}$ ($\tau=\SI{41}{\Gyr}$). Although essentially ruled out, the second model is included to understand better the effect on baryonic feedback of extreme cosmological excursions, which must necessarily be modelled in Monte Carlo analyses. For the same reason, we also extend the main FLAMINGO suite by running an additional $(\SI{1}{\Gpc})^3$ simulation with a large neutrino mass of $\sum m_\nu=\SI{0.48}{\eV}$, which is similarly ruled out by Planck (but see \citealt{divalentino22}). We refer the reader to Table~\ref{tab:sims} for an overview of the large simulations used in this paper.

\subsection{Additional simulations}\label{sec:additional_sims}

While the simulations listed in Table~\ref{tab:sims} have varying cosmological parameters, they are limited to specific scenarios or shifts in individual parameters. To mimic the type of parameter variations encountered in a Monte Carlo analysis, we ran 7 additional simulations with the same resolution but in a smaller volume (side length $L=\SI{200}{\Mpc}$) with cosmological parameters $(h,\Omega_\text{b}h^2,\Omega_\text{m}h^2,n_\mathrm{s},\sigma_8)$ sampled in a Latin hypercube with parameter ranges set by the $\pm3\sigma$ error bars around the best-fitting Planck TTTEEE + lowE model \citep{planck18}, except in the case of the power spectrum normalization, $\sigma_8$, for which we used the $\pm5\sigma$ range:
\begin{align}
    \begin{split}
    h&\in[0.657,\,0.6894],\\
    \Omega_\text{b} h^2&\in[0.021933,\,0.022833],\\
    \Omega_\text{m} h^2&\in[0.13984,\,0.14644],\\
    n_\mathrm{s}&\in[0.95345,\,0.97865],\\
    \sigma_8&\in[0.776,\,0.848].
    \end{split}\label{eq:cosmo_pars}
\end{align}
\noindent
To confirm an assumption of the model developed in Section~\ref{sec:results}, related to the scaling of the black hole mass, we ran another 4 simulations ($L=\SI{200}{\Mpc}$) identical to the fiducial FLAMINGO case but with the baryon fraction, $f_\mathrm{b}=\Omega_\mathrm{b}/(\Omega_\mathrm{b}+\Omega_\mathrm{c})$, varied by $\Delta f_\mathrm{b}/f_\mathrm{b}\in\{-4\%,-2\%,+2\%,+4\%\}$. These small simulations all used the fiducial subgrid physics model and a fixed neutrino mass of $\sum m_\nu=\SI{0.06}{\eV}$, modelled as one massive and two massless species. Gravity-only counterparts were also run for each model.

\section{The dependence of feedback on cosmology}\label{sec:results}

We begin our analysis by studying the cosmology dependence of the baryonic effect on the matter power spectrum in Section \ref{sec:global}, demonstrating the existence of non-factorizable corrections. The baryonic effect, which corresponds to the combined impact of the processes mentioned above (e.g. gas cooling, star formation, AGN and SN feedback) is measured by comparing the power spectrum in hydrodynamical simulations with that in dark matter only simulations. The essential role of the halo concentration in explaining the non-factorizable corrections is discussed in Section \ref{sec:properties}. We then construct a simple model for the non-factorizable correction to the power spectrum in Section~\ref{sec:towards}, before fitting it to the simulations in Section~\ref{sec:model}.

\subsection{Non-factorizable corrections to the matter power spectrum}\label{sec:global}

\begin{figure}
    \normalsize
    \centering
    \subfloat{
        \includegraphics{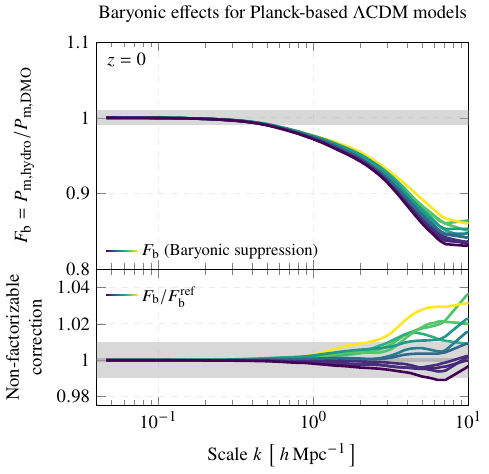}
    }
    \caption{The baryonic effect on the matter power spectrum, $P_\mathrm{m}$, in 11 $\Lambda$CDM models close to the best-fitting Planck model. The bottom panel shows that the non-factorizable corrections can be as large as $4-5\%$ even for small shifts in standard cosmological parameters. The solid lines correspond to the 5-bin central moving average of the data. The grey band represents a $1\%$ error. The colours indicate the average deviation from the baseline model.}
    \label{fig:separable_accuracy}
\end{figure}

We define the baryonic suppression of the matter power spectrum as the ratio
\begin{align}
    F_\text{b}(k)=\frac{P_\text{m}^\text{hydro}(k)}{P_\text{m}^\text{DMO}(k)},
\end{align}
\noindent
of the hydrodynamical matter power spectrum, $P_\text{m}^\text{hydro}(k)$, to the gravity-only matter power spectrum, $P_\text{m}^\text{DMO}(k)$. The top panel of Fig.~\ref{fig:separable_accuracy} shows this suppression at $z=0$ for a set of cosmological models that are close to the best-fitting Planck $\Lambda$CDM model \citep{planck18}. The results are based on the 11 $(\SI{0.2}{\Gpc})^3$ simulations described in Section~\ref{sec:additional_sims}. Despite the smaller volume of these simulations, the dependence of $F_\mathrm{b}$ on cosmology can be determined to within about $0.5\%$ on the scales of interest (see Appendix~\ref{sec:convergence}).

Let us consider first the general trend. The ratio equals unity on large scales up to about $k\approx\SIF{0.5\hub}{\per\Mpc}$ and then decreases to a dramatic minimum at $k\approx\SIF{10\hub}{\per\Mpc}$. This happens primarily because feedback from AGN expels gas from $10^{14}\si{M_\odot}$ halos, which lowers their contribution to the power spectrum at these scales \citep{vandaalen11,semboloni11,schneider19,debackere20}. Although we focus here on the scales $k\leq\SIF{10\hub}{\per\Mpc}$ most relevant for weak lensing observations, the suppression eventually turns into an enhancement of clustering for $k\geq\SIF{30\hub}{\per\Mpc}$ because gas cooling and star formation allows halos to contract, increasing the density on small scales. Precisely characterizing this is beyond the scope of the FLAMINGO project.

Assuming that the baryonic suppression is independent of cosmology, we may factorize the change in the power spectrum from one model (1) to another (2) as
\begin{align}
\frac{P^\text{hydro}_2}{P^\text{hydro}_1} &= \left(\frac{P^\text{DMO}_2}{P^\text{DMO}_1}\right)\left(\frac{P^\text{hydro}_2 / P^\text{DMO}_2}{P^\text{hydro}_1 / P^\text{DMO}_1}\right) \approx \frac{P^\text{DMO}_2}{P^\text{DMO}_1}.
\end{align}

\noindent
We refer to the second term in brackets as the \emph{non-factorizable correction} to this approximation. This factor is implicitly assumed to be 1 in works that ignore the cosmological coupling. The bottom panel of Fig.~\ref{fig:separable_accuracy} shows the correction for Planck-based variations in cosmology, assuming fixed astrophysical parameters and subgrid modelling. Varying the cosmological parameters by a few percent produces corrections on scales $k>\SIF{2\hub}{\per\Mpc}$ of up to $4-5\%$, which is relatively large compared to the total baryonic effect of $10-15\%$. Notice also that the deviations are mostly monotonic on these scales. This suggests that there is a systematic trend to be uncovered.

\begin{figure*}
    \normalsize
    \centering
    \subfloat{
        \includegraphics{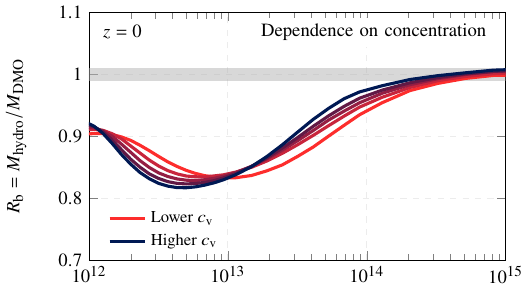}
    }
    \subfloat{
        \includegraphics{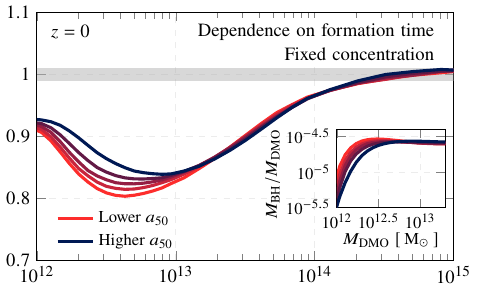}
    }~\\\vspace{-1em}
	\subfloat{
        \includegraphics{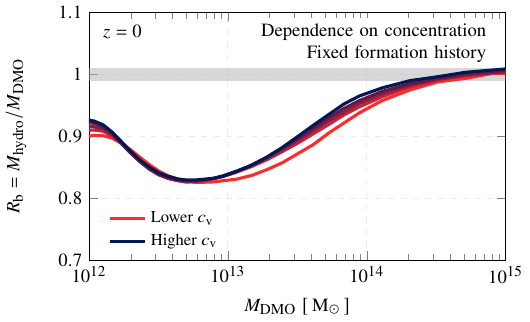}
    }
    \subfloat{
        \includegraphics{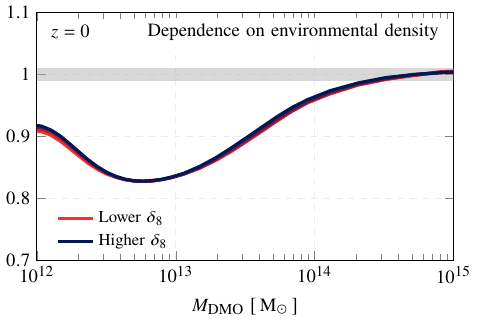}
    }
    \caption{The influence of different halo properties on the baryonic effect on halo mass, $R_\mathrm{b} = M_\text{hydro}/M_\text{DMO}$, comparing the masses of matched halos in the corresponding gravity-only and hydrodynamical simulations. We show the dependence of the mass ratio on secondary halo properties: concentration (top left), formation time at fixed concentration (top right), concentration at fixed formation history (bottom left), and environmental density (bottom right). The inset graph in the top-right panel shows how the ratio of black hole mass to gravity-only mass, $M_\text{BH}/M_\text{DMO}$, depends on mass and formation epoch. The colours indicate quintiles of the secondary halo property, calculated within each mass bin, with red indicating a lower value of that property and dark blue a higher value.}
    \label{fig:hydro_ratios}
\end{figure*}

\subsection{The role of halo concentration}\label{sec:properties}

The baryonic effect on the matter power spectrum can be understood as an aggregate of the effects on individual dark matter halos, integrated over a range of masses, since the stars and black holes that are responsible form inside halos. Let us therefore consider the baryonic effect on individual bound structures. The idea is that a change in cosmology alters the halo population, which subsequently leads to a change in the baryonic effect. We enumerated earlier a number of hypotheses for ways in which cosmology might affect feedback: by changing the structure of dark matter halos, their formation histories, or their environments.\footnote{Or their baryonic contents, but we will return to that crucial factor later.} We will argue that the first two mechanisms -- halo structure and formation history -- both play a role to some extent, but that for the range of scales most relevant for weak lensing observations, the first mechanism alone is important. Remarkably, we can neatly separate out these two mechanisms, depending on the mass of the halo. Finally, we will show that the influence of the environmental density around the halo is negligible.

To proceed, we match each halo in the hydrodynamical simulation with a halo in the gravity-only version of the same simulation by pairing the ten most strongly bound particles\footnote{Sorting halo particles according to their binding energies and using only a small number of most bound particles improves the quality of the matching.}, which allows us to determine the properties of the same halo with and without baryonic effects. This is done without varying the cosmological model, instead relying on the inherent scatter in halo properties in the large L2p8\_m9 simulations. The hydrodynamical simulation has 30 million halos with masses exceeding $\SIF{10^{12}}{M_\odot}$ and 0.2 million halos with masses exceeding $\SIF{10^{14}}{M_\odot}$, providing a large enough sample to make additional cuts based on secondary and tertiary halo properties. Our matching procedure results in a successful bijective pairing for $98.4\%$ of halos with $M_{200\text{c}}>\SIF{10^{12}}{M_\odot}$ and $99.2\%$ of halos with $M_{200\text{c}}>\SIF{10^{13}}{M_\odot}$.\footnote{Here, $M_{200\mathrm{c}}$ corresponds to the total mass contained in a spherical region with an average density equal to 200 times the critical density.} We are interested in the baryonic effect on the halo mass, for which we define the ratio
\begin{align}
R_\text{b}(M_\text{DMO}) = \frac{M_\text{hydro}}{M_\text{DMO}},
\end{align}

\noindent
using $M_{200\text{c}}$ masses. We exclusively consider central halos, ignoring satellites. The baryonic effect, $R_\text{b}$, depends sensitively on the gravity-only mass. We therefore split the sample into bins of $M_\text{DMO}$. Within each bin, we rank the halos according to a secondary halo property (see below) and compute $R_\text{b}$ for the five quintiles: the equal-sized groups with normalized rank between $[0,\,0.2]$ up to $[0.8,\,1]$. The mass bins are small enough that the correlation between the secondary property and halo mass is minimal. In each case, we compute the halo property from the gravity-only simulation and consider its role in determining the suppression of the halo mass in the hydrodynamical version, such that the direction of causality is unambiguous. We focus on the following properties:
\begin{enumerate}
    \item \emph{Maximum circular velocity, $V_\mathrm{max}$.}\quad The maximum circular velocity is defined in terms of the cumulative radial mass profile $M(\leq r)$, computed using all particles bound to the main subhalo, as $V_\text{max} = \max_{r>0} \sqrt{GM(\leq r)/r}$, where the radius $r$ is relative to the centre of potential.
	\item \emph{Concentration, $c_\mathrm{v}$.}\quad We use a velocity-based proxy for the halo concentration,
    \begin{align}
        c_\mathrm{v} = V_\text{max}/V_{200\text{c}},
    \end{align}
    
    \noindent
    using the maximum circular velocity, $V_\text{max}$, and virial velocity, $V_{200\mathrm{c}}=\sqrt{GM_{200\mathrm{c}}/R_{200\mathrm{c}}}$. For the usual definition of concentration in terms of the scale radius, $R_\mathrm{s}$, we have $c=R_{200\mathrm{c}}/R_\mathrm{s}\propto c_\mathrm{v}^2$ \citep{springel08,prada11}. This velocity-based definition of concentration remains sensible when the density profile cannot be determined reliably and can be computed directly from properties available in our halo catalogues.
	\item \emph{Formation epoch, $a_x$.}\quad The scale factor time, $a$, at which the halo first accreted $x\%$ of its present-day mass, computed by linearly interpolating between adjacent snapshots. We will consider both $a_{25}$ and $a_{50}$.
	\item \emph{Environmental density, $\delta_r$.}\quad We characterize the environment by computing the total mass, $M_r$, enclosed by halos within a radius of $r\,\si{\Mpc}$, excluding the halo itself. The environmental density, $\delta_r$, is then defined as $\delta_r = M_r/\langle M_r\rangle - 1$, where $\langle M_r\rangle$ is the average value for all halos in the sample. We will use $\delta_8$.
\end{enumerate}

\noindent
The resulting $R_\text{b}(M_\text{DMO})$ curves, split into quintiles of the secondary halo property, are shown in Fig.~\ref{fig:hydro_ratios}. Each curve shows the median of $R_\mathrm{b}$ in a given bin of mass and secondary property. First of all, we note that the general trend arises from the interplay between the depth of the gravitational potential well and the strength of stellar and AGN feedback at those masses \citep{cui14,velliscig14}. For the lowest mass shown, $\SIF{10^{12}}{M_\odot}$, AGN feedback is not important and the baryonic suppression of the mass is mainly driven by the outflow of gas due to supernovae. As the mass increases, AGN feedback becomes increasingly important and this drives the hydrodynamical mass down relative to the gravity-only mass, even as SN feedback becomes less effective. Finally, the ratio eventually approaches unity for the most massive halos beyond a few times $\SIF{10^{14}}{M_\odot}$, since even though gas is still ejected from massive galaxy clusters, particularly from satellites and progenitors of satellites, this is increasingly compensated by infalling gas \citep{mitchell20,mitchell22,wright24}.

Let us now consider the effect of the halo concentration, $c_\mathrm{v}$, in greater detail. This is shown in the top left panel of Fig.~\ref{fig:hydro_ratios}. There is a clear dependence of $R_\text{b}$ on concentration. For halos with dark matter masses between $10^{12}\si{M_\odot}$ and $10^{13}\si{M_\odot}$, the suppression is smaller for less concentrated halos, but the trend reverses for $M>10^{13}\si{M_\odot}$. There are different mechanisms at play in these two r\'egimes. Halo concentration is anti-correlated with formation epoch, older halos being more concentrated on average, as the cosmic matter density was higher at the time of their formation \citep[e.g.][]{navarro96b,navarro96,wechsler02,ludlow13}. In the low mass range, the dependence on concentration is due to the sensitivity of the black hole mass to the formation epoch of the halo. In the high mass range, the dependence on concentration is instead due to the binding energy of the halo rather than due to the formation history.

To see this, consider the dependence on the formation epoch shown in the top right panel for $a_{50}$. In this case, we additionally restrict the halo concentration, $c_\mathrm{v}$, to the $[40\%,\,60\%]$ interpercentile range to reduce the correlation between $c_\mathrm{v}$ and $a_{50}$. For halos $M<10^{13}\si{M_\odot}$, the suppression does depend on formation epoch. In this mass range, the masses of older halos are more strongly suppressed. This could be due to early galaxy mergers triggering rapid growth of the central supermassive black hole \citep{davies22} or simply because there has been more time for mergers and accretion \citep{matthee17}. There is no dependence on $a_{50}$, however, for halos beyond $10^{13}\si{M_\odot}$. This shows that the dependence on concentration seen in that mass range is not due to the formation epoch. To understand this, we show the black hole mass relative to the gravity-only mass, $M_\text{BH}/M_\text{DMO}$, in the inset graph. We see that halos with masses between $10^{12}\si{M_\odot}$ and $10^{13}\si{M_\odot}$ fall into a critical transition range where non-linear black hole growth is on the cusp of being triggered \citep{bower17,mcalpine18}, depending on the mass and formation epoch. At the high mass end, all black holes are self regulating, limiting their own growth by stopping the inflow of gas, and attain a more or less universal mass as a function of the dark matter mass and concentration, as might be expected from black hole scaling relations \citep{booth10,booth11}. From this point on, the dependence on formation epoch essentially disappears. If $M_\text{BH}/M_\text{DMO}$ is kept fixed, the dependence of $R_\text{b}$ on concentration becomes monotonic, with more concentrated halos suffering a smaller suppression (not shown). This confirms that the formation epoch plays a role in determining when the initial growth stage of the black hole occurs, but that the halo concentration determines the intensity of AGN feedback once the black holes become self regulating.

In the bottom left panel, we once again show the dependence on concentration, $c_\mathrm{v}$, but now controlling for the formation history. We do this by restricting both formation time proxies, $a_{25}$ and $a_{50}$, to their $[30\%,\,70\%]$ interpercentile ranges. In the low mass range around $\SI{5e12}{M_\odot}$, the dependence on concentration disappears. However, beyond $\SIF{10^{13}}{M_\odot}$, we recover the behaviour seen in the top left panel, with more concentrated halos experiencing a smaller baryonic suppression. We attribute this to the increased gravitational binding energies of the dark matter halos, trammelling the outflows driven by AGN.

A third possibility is that cosmology affects feedback by altering the halo environment. We study this possibility by determining the dependence of $R_\text{b}$ on the environmental density, $\delta_8$, defined in terms of the mass contained by halos within $\SI{8}{\Mpc}$. The bottom right panel of Fig.~\ref{fig:hydro_ratios} shows that this property is barely correlated with the baryonic suppression of halo mass. This may at first seem counterintuitive, given that halos with greater concentrations are found in denser regions \citep{avilareese05,wechsler06}. However, the environmental dependence of concentration is very weak compared to the scatter in concentrations at fixed mass \citep{maccio07}. The conclusion is the same for different definitions of $\delta_r$, such as $\delta_4$ or $\delta_{12}$, as well as definitions based on the total matter density. Hence, the large-scale environmental density does not appear to play a major role in regulating feedback.

Returning to the matter power spectrum, we note that the baryonic suppression on scales $\SIF{0.1\hub}{\per\Mpc}\leq k\leq\SIF{10\hub}{\per\Mpc}$ is mostly determined by halos with masses $10^{13}\si{M_\odot}<M<10^{14}\si{M_\odot}$ \citep[e.g.][]{semboloni11,debackere20,salcido23,vanloon24}. The results of this section indicate that the strength of baryonic feedback in this mass range is strongly correlated with the binding energy, as described by the halo concentration. Hence, a model of the non-factorizable corrections should first account for the change in halo concentration.

\subsection{A simple physical model}\label{sec:towards}

Let us now develop a physical model for the non-factorizable correction that also incorporates the dependence on the universal baryon fraction, $f_\mathrm{b} = \Omega_\text{b}/(\Omega_\text{b} + \Omega_\text{c})$. We focus on AGN feedback, which is the primary mechanism responsible for the baryonic suppression of the matter power spectrum on large scales, given the halo masses involved. The energy injected by the AGN is directly proportional to the black hole mass accretion rate:
\begin{align}
\dot{E}_\text{AGN} = \epsilon_\mathrm{r}\epsilon_\mathrm{f} \dot{m}_\text{accr}c^2,
\end{align}
\noindent
where $\epsilon_\mathrm{r}$ is the radiative efficiency of the black hole and $\epsilon_\mathrm{f}$ is the efficiency with which the energy is coupled to the gas. In the FLAMINGO simulations, $\epsilon_\mathrm{r}\epsilon_\mathrm{f}=0.015$ is fixed to reproduce the observed black hole scaling relations \citep{booth09}. The growth rate of the black hole mass is $\dot{M}_\text{BH}=(1-\epsilon_\mathrm{r})\dot{m}_\text{accr}$. Therefore, neglecting the small initial seed mass, the total energy injected can be related to the final black hole mass as
\begin{align}
E_\text{AGN} = \epsilon_\mathrm{r}\epsilon_\mathrm{f} \int\mathrm{d}t\,\dot{m}_\text{accr}c^2 = \frac{\epsilon_\mathrm{r}\epsilon_\mathrm{f}}{1-\epsilon_\mathrm{r}} M_\text{BH}c^2. \label{eq:Eagn_mbh}
\end{align}
\noindent
This energy is used to heat gas around the black hole in the centre of the galaxy. As a consequence, gas is ejected from the central region, thereby limiting further black hole growth. The gravitational binding energy of gas that once occupied the central region at some small radius $R$, but is now located at some larger radius $R'\gg R$, must decrease significantly. Let us assume that the energy injected by the AGN is used to lower the gravitational binding energy:
\begin{align}
E_\text{AGN} = -\Delta E_\text{grav} = E^\text{ini}_\text{grav} - E^\text{fin}_\text{grav} \approx E^\text{ini}_\text{grav}. \label{eq:Eagn}
\end{align}
\noindent
Consider the initial time before the rapid growth phase of the black hole was triggered and gas was ejected. Let $M$ be the total mass and $M_\text{b} = f_\text{b}M$ the baryonic mass contained in the central region with radius $R$. Then the initial gravitational binding energy is
\begin{align}
E^\text{ini}_\text{grav} = \frac{GMM_{\text{b}}}{R}.  \label{eq:Egravini}
\end{align}
\noindent
We will assume that baryons initially dominated the gravitational potential at the centre, such that $M\approx M_\text{b}$ and $E_\text{grav}\propto f_\text{b}^2$. On the other hand, if dark matter dominated the potential, $E_\text{grav}\propto f_\text{b}$. In general, we may write $E_\text{grav}\propto f_\text{b}^{n}$. Continuing with $n=2$, we find that the black hole mass should scale as
\begin{align}
M_\text{BH}c^2 \propto E_\text{AGN} \propto f_\mathrm{b}^{2} \frac{GM^2}{R}. \label{eq:mbh_scaling}
\end{align}
\noindent
Here, it is worth pausing and considering the implications so far. The idea of relating the black hole mass to the gravitational binding energy of the gas is not new \citep{silk98,ostriker05,booth10,zubovas16,bower17,oppenheimer18,chen20,davies20}. However, many previous models inferred that $M_\text{BH}\propto E_\text{grav} \propto f_\mathrm{b}$ with $n=1$ \citep[e.g.][]{zubovas16,chen20,davies20}, which would be the case if the gravitational potential were initially dominated by dark matter rather than baryons.

 \begin{figure}
     \normalsize
     \centering
     \subfloat{
         \includegraphics{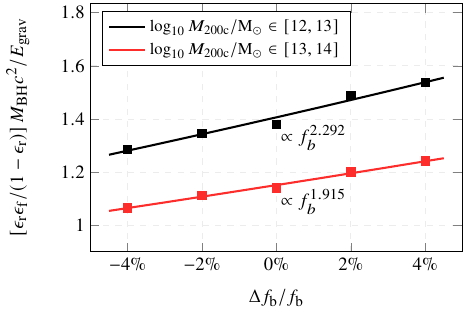}
     }
     \caption{The ratio of the black hole subgrid mass energy available for feedback, $\epsilon_\mathrm{r}\epsilon_\mathrm{f}/(1-\epsilon_\mathrm{r})M_\text{BH}c^2$, to the gravitational binding energy of the halo, $E_\text{grav}\equiv G M(\leq R)^2/R$, in a central sphere with $R=\SI{100}{\kpc}$, as a function of the universal baryon fraction, $f_\mathrm{b}$. We compute the median of this ratio in two mass bins for matched halos in five simulations varying only the universal baryon fraction, with binning based on the central $(\Delta f_\mathrm{b}=0)$ simulation. We find that the mass scales as $M_\text{BH}c^2\propto f_\mathrm{b}^2 E_\text{grav}$.}
     \label{fig:bh_fb_scaling}
 \end{figure}

We verified for the clusters in our simulations that the mass density is dominated by baryons at high redshift for $R\lesssim\SI{10}{\kpc}$, due to early star formation and gas cooling. This is in line with the expectation from higher resolution simulations \citep[e.g.][]{schaller15}. However, we caution that the density profile is not converged at these small radii at the fiducial FLAMINGO resolution. As an alternative way to verify our assumptions, we ran a number of simulations varying only the universal baryon fraction by $\Delta f_\mathrm{b}/f_\mathrm{b}\in\{-4\%,-2\%,+2\%,+4\%\}$, relative to the fiducial cosmology. We compute the ratio of the black hole mass energy available for feedback, $\epsilon_\mathrm{r}\epsilon_\mathrm{f}/(1-\epsilon_\mathrm{r}) M_\text{BH}c^2$, to the gravitational binding energy, $E_\text{grav}=GM^2/R$. We fix the radius of the central region at $R=\SI{100}{\kpc}$, which is resolved and corresponds approximately to the scale radius of $10^{13}\si{M_\odot}$ halos. The results are shown in Fig.~\ref{fig:bh_fb_scaling} for halos in two mass bins with $\log_{10}(M_{200\text{c}}/\si{M_\odot})\in[12,13]$ and $[13,14]$. While we emphasize that our assumptions are simplistic, ignoring for instance the dependence of radiative cooling losses on the baryon fraction, the proportionality \eqref{eq:mbh_scaling} is reproduced by the simulations. This suggests that it is the binding of baryons to baryons that determines the black hole mass in the FLAMINGO simulations.

Next, let us assume that the gas heated by the AGN and ejected from the central region has a velocity distribution $f(v/V_\mathrm{g})$, with scale parameter $V_\mathrm{g}$. We assume that
\begin{align}
V_\mathrm{g} = \sqrt{E_\text{AGN} / M_\text{b}} = \sqrt{f_\mathrm{b} GM/R}, \label{eq:Vg_def}
\end{align}
\noindent
where we used Eqs.~(\ref{eq:Eagn}--\ref{eq:mbh_scaling}). This is consistent with results from the EAGLE and IllustrisTNG simulations \citep{nelson19,mitchell20}, which show that black hole outflows have a velocity distribution that peaks at a characteristic velocity, $V_\mathrm{g}$, that increases with mass. The fraction of the gas at radius $r$ that escapes the halo corresponds to the fraction with a velocity that exceeds the escape velocity,
\begin{align}
V_\text{esc}\approx V_\text{max}\sqrt{\frac{2\log(1+2r/R_\text{max})}{r/R_\text{max}}}\approx 2V_\text{max},
\end{align}
\noindent
where $R_\text{max}$ is the radius where the circular velocity is $V_\text{max}$, the expression holds for an NFW profile \citep{navarro96,klypin97}, and we assumed that $r\ll R_\text{max}$ in the second step. The factor multiplying $V_\text{max}$, given by the square root, is not very sensitive to $r$. Hence, the escape fraction is approximately
\begin{align}
f_\text{esc} &= \int_{2V_\text{max}}^\infty \mathrm{d}v' f(v'/V_\mathrm{g}) = f_\text{esc}(\xi),
\end{align}
\noindent
where upon evaluation, using dimensional arguments, we write the integral as an arbitrary function of the ratio
\begin{align}
\xi \equiv \frac{V_\mathrm{g}}{V_\text{max}} = \frac{\sqrt{f_\mathrm{b}GM/R}}{V_\text{max}} = \frac{\sqrt{f_\mathrm{b}}}{c_\mathrm{v}}.
\end{align}
\noindent
Notice that the non-linear scaling of the black hole mass, $M_\text{BH}\propto f_\mathrm{b}^n$, was important, since $f_\mathrm{b}$ would have dropped out if $n=1$, as assumed by earlier works. In general, repeating the calculation for arbitrary $n$, we predict that $f_{\text{esc}}$ should be a function of $f_\mathrm{b}^{(n-1)/2}c^{-1}_v$. The scaling $M_\text{BH}\propto f_\mathrm{b}^2$ should be confirmed with other simulations and feedback models. Moreover, the scaling should be confronted with observational lines of evidence. One possibility is to make use of the fact that baryons are not exact tracers of the dark matter even at early times \citep{angulo13}, giving rise to spatial variations in the baryon fraction and therefore perhaps spatial variations in the strength of feedback.

\begin{figure}
    \normalsize
    \centering
    \subfloat{
        \includegraphics{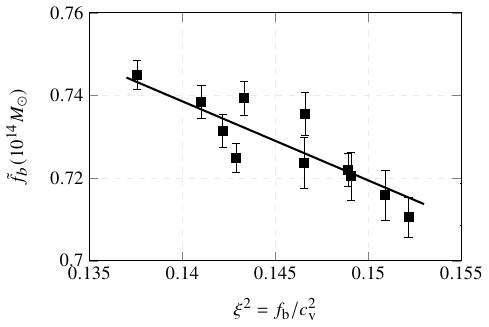}
    }
    \caption{A fit of renormalized baryon fractions, $\tilde{f}_\mathrm{b}=f_{\mathrm{b},200\mathrm{c}}/f_\mathrm{b}$, where $f_{\mathrm{b},200\mathrm{c}}$ is the baryon fraction within $R_{200\mathrm{c}}$ for matched halos with a mass of $M_{200\mathrm{c}}=10^{14}\si{M_\odot}$ and $f_\mathrm{b}$ is the universal baryon fraction, as a function of the cosmological parameter combination $\xi^2=f_\mathrm{b}/c_\mathrm{v}^2$. The baryon fractions are evaluated at the pivot mass, $10^{14}\si{M_\odot}$, using the power law fits of Eq.~\eqref{eq:fb_power_law_fit}.} 
    \label{fig:fbar_clusters}
\end{figure}

\begin{figure*}
    \normalsize
    \centering
    \subfloat{
        \includegraphics{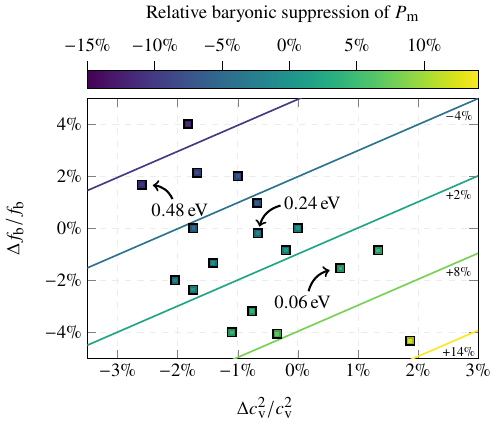}
    }
    \subfloat{
        \includegraphics{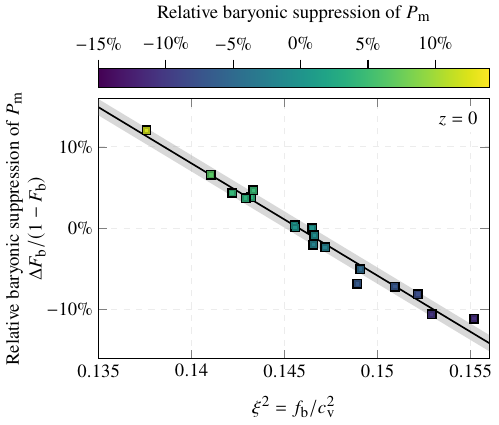}
    }\vspace{-0.5em}
    \caption{The non-factorizable correction to the power spectrum as a function of the changes in baryon fraction, $f_\mathrm{b}$, and gravity-only halo concentration, $c_\mathrm{v}$. The points are based on FLAMINGO simulations with side lengths $L=\SI{1}{\Gpc}$ and $L=\SI{0.2}{\Gpc}$, shown relative to the fiducial L1\_m9 model at $(\Delta c^2_\mathrm{v},\Delta f_\mathrm{b})=(0,0)$. The arrows on the left indicate three models that only differ in the assumed neutrino mass sum of $\sum m_\nu=\SI{0.06}{\eV},$ $\SI{0.24}{\eV}$ or $\SI{0.48}{\eV}$. The right-hand panel shows that Eq.~\eqref{eq:phys_model2} provides an excellent fit, explaining the correction to the power spectrum in terms of $\xi^2=f_\mathrm{b}/c_\mathrm{v}^2$. The grey band represents a $\pm1\%$ error in the relative suppression of the power spectrum, implying an even smaller error in $P_\text{m}(k)$.}
    \label{fig:pk_vs_c}
\end{figure*}

Let us now see how the full model compares against the simulations. For each of the cosmological model variations described in Section~\ref{sec:additional_sims}, we take the median of the concentration, $c_\mathrm{v} = V_\text{max}/V_{200\mathrm{c}}$, for halos with masses $M_{200\mathrm{c}}$ in the range $10^{14\pm0.25}\si{M_\odot}$ from the gravity-only simulations. We compute the universal baryon fraction, $f_\mathrm{b} = \Omega_\text{b}/(\Omega_\text{b} + \Omega_\text{c})$, directly from the cosmological parameters. These two quantities give the cosmological parameter combination $\xi = \sqrt{f_\mathrm{b}}/c_\mathrm{v}$. Determining the escape fractions is more complicated because of the limited volume and the stochasticity of the hydrodynamical simulations. To get a statistically significant relationship, we again rely on pairs of halos matched between the fiducial cosmology and the variations. For each halo, we compute renormalized baryon fractions,
\begin{align}
    \tilde{f}_\mathrm{b} = \frac{f_{\mathrm{b},200\mathrm{c}}}{f_\mathrm{b}},
\end{align}

\noindent
where $f_{\mathrm{b},200\mathrm{c}}$ is the baryon fraction within $R_{200\mathrm{c}}$ and $f_\mathrm{b}$ is the universal baryon fraction. In the mass range of interest, $10^{13.25}<M_{200\mathrm{c}}/\si{M_\odot}<10^{14.5}$, we fit a power-law relationship 
\begin{align}
    \tilde{f}_\mathrm{b}(M_{200\mathrm{c}}) = a \left(\frac{M_{200\mathrm{c}}}{10^{14}\si{M_\odot}}\right)^b, \label{eq:fb_power_law_fit}
\end{align}

\noindent
in terms of the free parameters $a$ and $b$, and evaluate $\tilde{f}_\mathrm{b}$ at the pivot mass, $10^{14}\si{M_\odot}$. Plots of these fits are shown in Fig.~\ref{fig:fbar_clusters_mass_fits} in Appendix \ref{sec:power_law_fits}. In Fig.~\ref{fig:fbar_clusters}, we show a fit of $\tilde{f}_\mathrm{b}$ in terms of $\xi^2$,
\begin{align}
    \tilde{f}_\mathrm{b} = 1 - f_\text{esc} = c + d \xi^2, \label{eq:ftilde}
\end{align}

\noindent
where $c$ and $d$ are free parameters. We find that $d<0$. Given that $\xi^2=f_\mathrm{b}/c_\mathrm{v}^2$, this is consistent with the expectation that the escape fraction increases with the universal baryon fraction, $f_\mathrm{b}$, and decreases with the binding energies of halos. In the next section, we will use this model to predict the non-factorizable corrections to the matter power spectrum.

\subsection{Fitting function for the non-factorizable corrections to the matter power spectrum}\label{sec:model}

The baryon fraction in groups can be related to the suppression of the matter power spectrum \citep[e.g.][]{semboloni13,debackere20,vandaalen20,salcido23,vanloon24}. At any fixed wavenumber $k$, the suppression of the power spectrum is well described by an exponential fit of the form:
\begin{align}
F_\mathrm{b} &= \frac{P^\text{hydro}_\text{m}}{P_\text{m}^\text{DMO}} = 1 - \exp\big(e\tilde{f}_\mathrm{b}+g\big),
\end{align}
\noindent
with the free parameters $e$ and $g$.\footnote{This does not imply that two parameters are needed per $k$-bin. On large scales, $F_\mathrm{b}(k,\tilde{f}_\mathrm{b})$ can be fitted with $5$ free parameters \citep{vandaalen20}.} Combining this with Eq.~\eqref{eq:ftilde}, we find that the suppression of the power spectrum should depend on cosmology as
\begin{align}
F_\mathrm{b} &= 1 - \exp(\alpha\xi^2+\beta) =  1 - \exp\left(\alpha \frac{f_\mathrm{b}}{c_\mathrm{v}^2}+\beta\right), \label{eq:phys_model}
\end{align}
\noindent
where $\alpha=de$ and $\beta = g+ce$. Going forward, we will simply consider $\alpha$ as a free parameter and fit the model to the power spectra in the simulations, expanding Eq.~\eqref{eq:phys_model} to first order and writing the change in baryonic suppression relative to the fiducial case as
\begin{align}
\frac{\Delta F_\mathrm{b}}{1-F_\mathrm{b}} &= -\alpha\Delta(\xi^2) \label{eq:phys_model2},
\end{align}

\noindent
Here, $\Delta F_\mathrm{b} = F_\mathrm{b} - F^\text{fid}_\mathrm{b}$, such that $\Delta F_\mathrm{b} > 0$ corresponds to weaker feedback and therefore greater $P(k)$. For each model variation, we compute the average baryonic suppression of the power spectrum, $F_\mathrm{b}$, in the range between $\SIF{0.1\hub}{\per\Mpc}\leq k \leq \SIF{10\hub}{\per\Mpc}$ at $z=0$. The median concentration, $c_\mathrm{v}$, and universal baryon fraction, $f_\mathrm{b}$, are determined as in the previous section. Fitting model~\eqref{eq:phys_model2} to the simulations yields $\alpha = 13.8\pm0.6$.

The left panel of Fig.~\ref{fig:pk_vs_c} shows the relative change in the baryonic suppression, $\Delta F_\mathrm{b} / (1-F_\mathrm{b})$, in the plane of $f_\mathrm{b}$ and $c_\mathrm{v}$. The contour lines indicate the level sets of $\Delta F_\mathrm{b}$. As expected, an increase in the universal baryon fraction leads to a greater baryonic suppression, while an increase in the halo concentration leads to a smaller suppression. From the slope of the contour lines, we see that the logarithmic slope of $F_\mathrm{b}$ with respect to $f_\mathrm{b}$ is approximately equal to the logarithmic slope with respect to $1/c^2_\mathrm{v}$, consistent with Eq.~\eqref{eq:phys_model}. Three simulations, which differ only in the assumed neutrino mass sum of $\sum m_\nu=\SI{0.06}{\eV}$, $\SI{0.24}{\eV}$, and  $\SI{0.48}{\eV}$, are marked in Fig.~\ref{fig:pk_vs_c}. The vector between these models makes a large angle with the contour lines, implying a significant change in baryonic suppression. The non-factorizable correction between these models is large for two reasons. First of all, halos are less concentrated in the large neutrino mass cosmology because the matter density is lower when halos collapse. Secondly, more gas is available for halos of a given dark matter mass, because neutrinos cluster less effectively on small scales, such that halos are primarily composed of CDM and baryons, of which baryons make up a larger fraction. Both changes increase the potency of feedback. This explains why, among the original 5 cosmology variations in the FLAMINGO suite \citep{schaye23}, the non-factorizable correction between these models is largest.

Much the same is true for the decaying dark matter simulations, but to an even greater extent. For these models, the baryon fraction increases over time as dark matter decays. To interpret the suppression of the power spectrum in terms of $\xi^2=f_\mathrm{b}/c_\mathrm{v}^2$, we choose to evaluate $f_\mathrm{b}$ at $z=0$, since we do the same for $c_\mathrm{v}$. Comparing the Planck simulations with decay rates of $\Gamma=\SI{12}{\km\per\s\per\Mpc}$ and $\SI{24}{\km\per\s\per\Mpc}$ to the fiducial cosmology without decays, we find changes in the baryon fraction of $\Delta f_\mathrm{b}/f_\mathrm{b} = 15\%$ and $30\%$, respectively, while the square of the median halo concentration changes by $\Delta c^2_\mathrm{v}/c^2_\mathrm{v}=-6\%$ and $-11\%$, respectively. Given that these points are far outside the parameter ranges of Fig.~\ref{fig:pk_vs_c}, we provide a separate plot with the decaying dark matter simulations in Fig.~\ref{fig:suppression_with_dcdm} in Appendix~\ref{sec:power_law_fits}. As expected, the baryonic suppression in these models is far stronger. This will be discussed further in the next section and in Appendix~\ref{sec:power_law_fits}

The right-hand panel of Fig.~\ref{fig:pk_vs_c} shows that Eq.~\eqref{eq:phys_model2} provides an excellent fit, giving the relative baryonic suppression, $\Delta F_\mathrm{b}/(1-F_\mathrm{b})$, as a function of $\xi^2=f_\mathrm{b}/c_\mathrm{v}^2$ to a relative accuracy of about $1\%$, implying an even greater absolute accuracy in $P_\text{m}$. Although we focus on the suppression at $z=0$ in this paper, the same general trends with respect to $f_\mathrm{b}$ and $c_\mathrm{v}$ remain true at higher redshifts. We will briefly revisit the time dependence in Section~\ref{section:s8_degeneracies}, but leave a detailed treatment for future work.

\section{Cosmological implications}\label{sec:application}

One issue that has received significant attention in recent years is the tension between different measurements of the $S_8=(\Omega_\text{m}/0.3)^{1/2}\,\sigma_8$ parameter. The values obtained from various large-scale structure probes have been found to be lower than the $\Lambda$CDM expectation based on measurements of the CMB. In particular, weak lensing surveys such as the Kilo-Degree Survey (KiDS; \citealt{kuijken15}) and Dark Energy Survey (DES; \citealt{abbott16}) have reported measurements that are $5\sim10\%$ lower than that of Planck \citep{planck18} with a statistical significance of $2\sim3\sigma$ \citep{amon22,asgari22,abbott23}. It has long been recognized that baryonic processes, in particular feedback from AGN, are relevant for the interpretation of weak lensing observations \citep[e.g.][]{semboloni11,vandaalen11}. As such, large-scale structure analyses usually take the impact of baryons into account, either by marginalizing over some parametrization of feedback or by excluding small-scale clustering where the impact of baryons is thought to be largest. However, \citet{amon22b} and \citet{preston23} have recently argued that lensing- and CMB-based measurements of $S_8$ can be reconciled by adopting a baryonic feedback model that is significantly stronger than what is predicted by hydrodynamical simulations and X-ray observations of clusters. That such a solution is not feasible is confirmed by results from the new FLAMINGO suite \citep{mccarthy23}, which suggest that baryonic effects are too small even when allowing for large uncertainties in the observed cluster gas fractions and theoretical modelling. While we focus on weak lensing in this paper, they also showed that other probes, such as the thermal Sunyaev-Zel’dovich (tSZ) effect, exhibit large tensions not reconciliable with baryonic feedback alone.

Setting aside the impact of baryons, the $S_8$ tension could also be interpreted as a hint of new physics that suppresses the growth of cosmic structure at late times. Such a modification of the concordance model must simultaneously be large enough to reconcile large differences in power spectra at low redshifts, without distorting the tightly constrained expansion history or the clustering at higher redshifts and larger scales effectively probed by CMB lensing. Adding to the difficulty is the fact that solutions to the $S_8$ tension often exacerbate the $H_0$ tension and vice versa \citep[e.g.][]{pandey20,vagnozzi23}. Given the coupling between baryonic feedback and cosmology discussed in this paper, it is worth considering whether a combination of baryonic and non-baryonic suppression mechanisms could work together, obviating the need for extreme versions of either mechanism alone. This will be discussed in the following sections. First, in Section~\ref{section:s8_ranges}, we will determine the cosmological parameter ranges within which a feedback model calibrated assuming a fiducial Planck cosmology can be reliably applied. Then, in Section~\ref{section:s8_extensions}, we will consider the implications for extensions of $\Lambda$CDM aimed at resolving the $S_8$ tension. Finally, we will look at the extent to which cosmological and astrophysical parameters are degenerate in Section~\ref{section:s8_degeneracies}.

\subsection{Parameter dependence}\label{section:s8_ranges}

To quantify the variations in baryonic feedback that accompany variations in cosmology, we must first express Eq.~\eqref{eq:phys_model2} in terms of standard cosmological parameters. Using the $\Lambda$CDM simulations described in Section~\ref{sec:additional_sims}, we obtain a fit for our velocity-based definition of halo concentration, for halos with masses $M_{200\mathrm{c}}=10^{14\pm0.25}\si{M_\odot}$, in terms of $\Omega_\text{m}$ and $\sigma_8$,
\begin{align}
    c_\mathrm{v} = 1.24 (\Omega_\text{m}\sigma_8)^{1/8}, \label{eq:concentration_fit}
\end{align}
\noindent
which is in line with the expectation that cosmological model variations that reduce the strength of clustering lead to less concentrated halos, although the dependence is relatively weak. We show the scatter around this relationship in Fig.~\ref{fig:c_tilde_fit} in Appendix~\ref{sec:power_law_fits}. Substituting this expression into Eq.~\eqref{eq:phys_model2}, we obtain
\begin{align}
    \frac{\Delta F_\mathrm{b}}{1-F_\mathrm{b}} &= -\alpha' \Delta\left[ \frac{\Omega_\text{b}}{\Omega_\text{b}+\Omega_\text{c}}(\Omega_\text{m}\sigma_8)^{-1/4}\right], \label{eq:phys_model_cosmo}
\end{align}
\noindent
where $\alpha'=0.65\alpha=8.98$. This expression gives the relative change in the baryonic suppression of the matter power spectrum, averaged over the range $\SIF{0.1\hub}{\per\Mpc}<k<\SIF{10\hub}{\per\Mpc}$ at $z=0$. To simplify matters further, we observe that external constraints on the baryon density, $\Omega_\text{b}h^2$, from Big Bang Nucleosynthesis \citep{adelberger11,cooke18} are much stronger than external constraints on the dark matter density, $\Omega_\text{c}h^2$, justifying a much stronger prior on $\Omega_\text{b}h^2$. It follows that the baryon fraction, $f_\mathrm{b}=\Omega_\text{b}/(\Omega_\text{b}+\Omega_\text{c})$, is strongly anti-correlated with the total matter density, $\Omega_\text{m}$, in cosmic shear analyses (as it is for Planck). Computing the marginal expectation value, $f_\mathrm{b}(\Omega_\text{m})$, from the KiDS + BOSS + 2dFLenS $3\times 2$pt chains \citep{heymans21}, we can express $\Delta F_\mathrm{b}$ entirely in terms of $\Omega_\text{m}$ and $\sigma_8$. We obtain very similar results when using the marginal expectation $f_\mathrm{b}(\Omega_\text{m})$ from the Planck TTTEEE + lowE chains \citep{planck18}.

\begin{figure}
    \centering
    \includegraphics{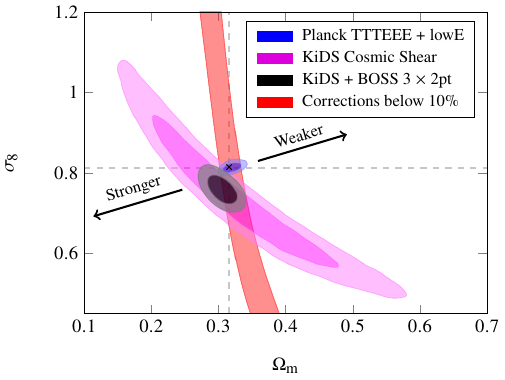}\vspace{-0.5em}
    \caption{The cosmological parameters for which the baryonic suppression of matter clustering is similar to that in the best-fitting Planck model (red). This corresponds to the range where absolute errors to the matter power spectrum are $\lesssim 1\%$, or more precisely where relative non-factorizable corrections, $\Delta F_\mathrm{b}/(1-F_\mathrm{b})$, are below $10\%$ on scales $\SIF{0.1\hub}{\per\Mpc}\leq k\leq\SIF{10\hub}{\per\Mpc}$ at $z=0$. The arrows indicate where baryonic feedback is at least $10\%$ stronger than in the Planck model (lower $\sigma_8$ and $\Omega_\text{m}$) and weaker (vice versa). This is to be compared with the marginal posteriors from KiDS-1000 (purple), KiDS + BOSS + 2dFLenS (black), and Planck TTTEEE + lowE (blue).}
    \label{fig:contours}
\end{figure}

\begin{figure*}
    \normalsize
    \centering
    \subfloat{
        \includegraphics{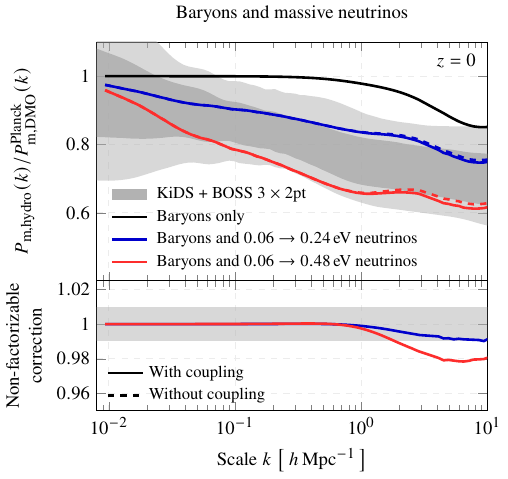}
    }
    \subfloat{
        \includegraphics{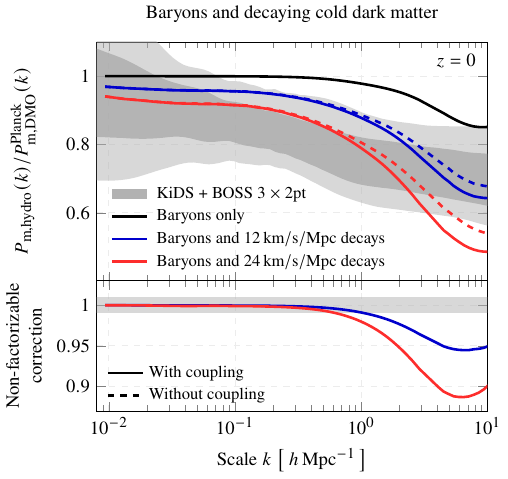}
    } \vspace{-0.5em}
    \caption{Matter power spectra relative to the spectrum for a Planck-based $\Lambda$CDM cosmology \citep{planck18} without baryonic effects. The grey contours in the top panel show the $68\%$ and $95\%$ ranges of $\Lambda$CDM spectra obtained from the $3\times2$pt KiDS + BOSS + 2dFLenS analysis \citep{heymans21}, which include the effects of baryonic feedback by marginalizing over a halo model-based feedback parameter (\citealt{mead15}; see Section~\ref{section:approximations}). The left panel shows the effects of baryons (black) and the effects of baryons combined with changing the neutrino mass from $\sum m_\nu=\SI{0.06}{\eV}$ to $\SI{0.24}{\eV}$ (blue) or $\SI{0.48}{\eV}$ (red). The solid lines represent the simulation results, while the dashed lines show the results without accounting for the cosmology dependence of baryonic feedback. The right panel shows the same for decaying cold dark matter (DCDM) models with decay rates of $\Gamma = \SI{12}{\km/\s/\Mpc}$ (blue) and $\SI{24}{\km/\s/\Mpc}$ (red). The bottom panels show the non-factorizable corrections, i.e. the ratios of the solid and dashed curves in the top panels, with grey bars representing a $1\%$ error.}
    \label{fig:separable_accuracy_dcdm}
\end{figure*}

Having eliminated $\Omega_\text{b}$, we compute the parameter ranges around the best-fitting Planck model, marked by a cross in Fig.~\ref{fig:contours}, for which the relative corrections, $\Delta F_\mathrm{b}/(1-F_\mathrm{b})$, are below $10\%$. The result is shown by the red contours in the figure. They are slightly tilted from vertical, reflecting the importance of the baryon fraction, $f_\mathrm{b}$, while the dependence on $\sigma_8$ through the halo concentration \eqref{eq:concentration_fit} is weaker. The importance of the baryon fraction is not surprising and has been noted previously \citep{schneider20,vandaalen20,arico21,mead21}. The secondary dependence on $\sigma_8$ is also consistent with \citet{arico21}, who used a baryonification algorithm to explore the dependence of feedback on cosmology\footnote{In the baryonification algorithm, feedback is accounted for by displacing particles according to a recipe with a number of parameters. In \citet{arico21}, the cosmological coupling is analysed with parameters tuned to hydrodynamical simulations. Hence, the conclusions are not entirely independent.}. Also shown in Fig.~\ref{fig:contours} are the constraints from Planck temperature and polarization data in blue, the KiDS-1000 weak lensing constraints in purple, and the KiDS + BOSS + 2dFLenS $3\times2$pt constraints in black. The figure shows that relative non-factorizable corrections are mostly below $10\%$ (which implies absolute corrections to the power spectrum below $1\%$) if one restricts to the $68\%$ Planck constraints. However, it is not possible to sample a parameter space that covers both the constraints from Planck and large-scale structure surveys like KiDS with $1\%$ absolute precision unless non-factorizable corrections are taken into account.

Another interesting observation is that baryonic effects are enhanced for lower $\sigma_8$ (decreasing the binding energies of halos) and lower $\Omega_\text{m}$ (both decreasing binding energies and increasing the baryon fraction). This is precisely the direction of the tension. Hence, new physics that involves lowering the density of matter or the amplitude of its clustering at late times also tends to increase the strength of feedback, and possibly more so than variations in standard cosmological parameters. We will explore this idea in the next section.

\subsection{Model building}\label{section:s8_extensions}

Within the $\Lambda$CDM model, a smaller lensing signal translates into a preference for reduced matter clustering. Although baryonic feedback does produce such an effect, the FLAMINGO model predicts that this is too small to reconcile all observations \citep{mccarthy23}. One way to illustrate this is to show matter power spectra relative to the power spectrum for a ``CMB cosmology'' without baryonic feedback. We do just that in Fig.~\ref{fig:separable_accuracy_dcdm}. The grey contours represent the $68\%$ and $95\%$ intervals of $\Lambda$CDM power spectra allowed by the $3\times2$pt KiDS + BOSS + 2dFLenS analysis \citep{heymans21}, including the effects of baryonic feedback (parametrized with the \textsc{hmcode} method of \citealt{mead15}, see Section~\ref{section:approximations}), relative to the gravity-only FLAMINGO simulation with the best-fitting values from Planck \citep{planck18}. This is computed by marginalizing over all cosmological, baryonic feedback, and nuisance parameters taken from the public chains. The black line shows the effect of turning on baryonic feedback in the Planck cosmology, assuming  the fiducial FLAMINGO parameters. Consistent with the headline result of a $\sim3\sigma$ tension in $S_8$ reported in \citet{heymans21}, the discrepancy still exceeds $2\sigma$ on scales $\SIF{0.1\hub}{\per\Mpc}<k<\SIF{10\hub}{\per\Mpc}$. It is worth noting that the KiDS + BOSS + 2dFLenS results plotted in Fig.~\ref{fig:separable_accuracy_dcdm} are more extreme than those of DES-Y3 \citep{abbott22,abbott23}, which are intermediate between KiDS and Planck.

One way to reconcile weak lensing and CMB observations is by introducing new physics that slows the growth of structure at late times. Two well-motivated extensions of the Standard Model that have been considered in this context are massive neutrinos \citep{wyman14,battye14,mccarthy18} and decaying dark matter \citep{aoyama14,berezhiani15,enqvist15,pandey20,abellan21,tanimura23}. Both models slow the growth of structure and increase the density of baryons relative to the matter that clusters efficiently. Based on the preceding discussion and that in Section~\ref{sec:model}, we conclude that both extensions will therefore also enhance the strength of baryonic feedback, further boosting the overall suppression of matter clustering.

\begin{figure*}
    \normalsize
    \centering
    \subfloat{
        \includegraphics{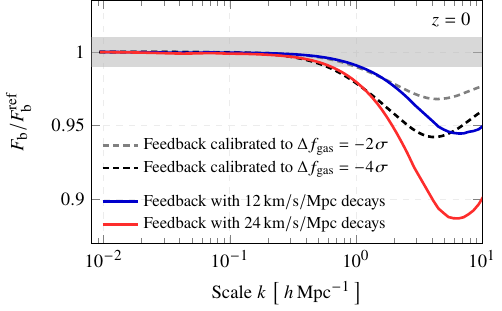}
    }
    \subfloat{
        \includegraphics{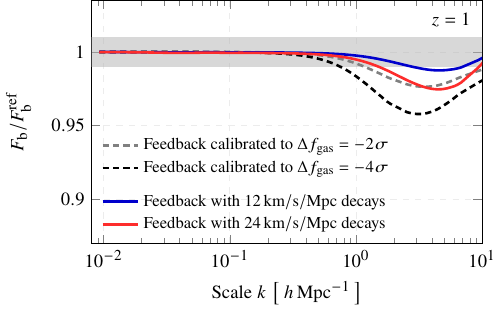}
    } \vspace{-0.5em}
    \caption{Ratios of the baryonic suppression of the power spectrum, relative to the fiducial case without decaying dark matter and with feedback parameters calibrated to observed cluster gas fractions, for models with different dark matter decay rates but equal subgrid parameters (blue and red) and for models without decaying dark matter but subgrid parameters calibrated to systematically lower cluster gas fractions (dashed grey and black). The two types of parameters have different time- and scale-dependent effects. The cosmological coupling becomes more important on small scales ($k\geq\SIF{1\hub}{\per\Mpc}$) and at late times $(z<1)$.}
    \label{fig:dcdm_agn_variations}
\end{figure*}

To see this more clearly, we show the effect of increasing the neutrino mass from $\sum m_\nu=\SI{0.06}{\eV}$ for the fiducial Planck cosmology to $\sum m_\nu=\SI{0.24}{\eV}$ (blue) or $\SI{0.48}{\eV}$ (red) on the left-hand side of Fig.~\ref{fig:separable_accuracy_dcdm}. The dashed lines indicate the effect that would result if baryonic feedback were independent of cosmological parameters, while the solid lines indicate the results obtained from the simulations. The bottom panel shows that the non-factorizable corrections are modest, with effects of $1\%-2\%$ for $k>\SIF{2\hub}{\per\Mpc}$, relative to the case without cosmological dependence, while also confirming that the suppression is indeed greater than in the case with minimal neutrino masses, $\sum m_\nu=\SI{0.06}{\eV}$. This agrees with the prior findings of \citet{mummery17} obtained with the BAHAMAS simulations. It also explicitly confirms the finding in Section~\ref{section:s8_ranges} that non-factorizable corrections exceed $1\%$ when moving from the best-fitting Planck model to a model preferred by large-scale structure surveys. It is striking to see that the $0.24-\SI{0.48}{\eV}$ simulations straddle the $1\sigma$ constraints from KiDS + BOSS + 2dFLenS. This is consistent with earlier studies attempting to reconcile lensing and CMB observations with neutrinos \citep{wyman14,battye14,mccarthy18}. However, such a solution is strongly challenged by geometric constraints obtained from the combination of CMB and BAO data \citep[e.g.][]{vagnozzi17,planck18,brieden22,tristram23}\footnote{In fact, the Planck$\nu$0.24Fix simulation shown in Fig.~\ref{fig:separable_accuracy_dcdm} is also ruled out by Planck alone, since the remaining cosmological parameters are not adjusted.}, the Lyman-$\alpha$ forest \citep{palanque20}, and to a much lesser extent by CMB lensing \citep{planck18}. 

For decaying dark matter, the effect on baryonic feedback is much stronger. This is shown on the right-hand side of Fig.~\ref{fig:separable_accuracy_dcdm}, for decay rates of $\Gamma=\SI{12}{\km/\s/\Mpc}$ (blue) and $\SI{24}{\km/\s/\Mpc}$ (red). Dashed lines again indicate the predictions without dependence on cosmology, while solid lines show the simulated results. The non-factorizable corrections are now $5-10\%$ on non-linear scales, essentially doubling the strength of feedback in the most extreme case relative to the model without decaying dark matter. This shows that the assumption made in \cite{hubert21} that baryonic feedback is not affected by dark matter decay is not a good approximation. It is interesting to note that decaying dark matter and warm dark matter are fundamentally different in this regard, since the factorizability approximation is more accurate for warm dark matter \citep{parimbelli21}. Compared to the neutrino models, the scale dependence of the combined baryonic and non-baryonic effect is less consistent with the $3\times2$pt constraints, although the models do provide a better fit than the Planck model with baryons alone. However, as in the case of massive neutrinos, geometric considerations also seem to rule out this solution \citep[e.g][]{audren14,aubourg15}. The non-factorizable corrections may also lead to violations of astrophysical constraints, such as cluster gas fractions. If this is the case, it could be another reason to exclude decaying dark matter as a viable solution. However, as mentioned in the introduction, turning baryonic feedback into a cosmological probe is only possible if the degeneracy between astrophysical and cosmological parameters could be broken. We will return to this issue shortly.

The decaying dark matter model studied in this paper is perhaps only the simplest, so it is worth considering other types of behaviour. Whereas all dark matter decays at a constant and universal rate in this model, the rate may depend on the density and velocity dispersion of dark matter, as in certain models with annihilating dark matter \citep{choquette16}. For such models, the effect would be most pronounced in the centres of massive halos, which we speculate would lead to even greater non-factorizable corrections, given the importance of the central density of groups and clusters for AGN feedback. Two-body dark matter decay, in which a less massive daughter dark matter particle receives a small recoil velocity, would heat or disrupt dark matter halos and lower their concentrations \citep{peter10,peter10b,cheng15,abellan20}, likely further enhancing baryonic feedback. On the other hand, interactions between dark matter and dark energy may either boost or suppress the halo concentration \citep{li11,baldi17,an19,liu21}, with uncertain effects on feedback.

\subsection{Degeneracies}\label{section:s8_degeneracies}

Given these findings, a further question is whether the cosmological dependence of baryonic feedback is degenerate with the astrophysical parameters of the model, which were kept fixed in all cosmological model variations considered so far.\footnote{We stress the difference between the factorizability of two processes (i.e. that they can be treated independently) and their degeneracy (i.e. that they have identical effects on some obserable).} There is some indication that this may not be the case. In the left panel of Fig.~\ref{fig:dcdm_agn_variations}, we show the change in baryonic suppression (the non-factorizable correction) at $z=0$ when setting the dark matter decay rate to $\Gamma=\SI{12}{\km/\s/\Mpc}$ (blue) or $\SI{24}{\km/\s/\Mpc}$ (red), whilst keeping the feedback parameters fixed, together with two variations in subgrid model at fixed cosmology. The two astrophysics variations, listed as L1\_m9\_$f_\text{gas}\!-2\sigma$ and L1\_m9\_$f_\text{gas}\!-4\sigma$ in Table~\ref{tab:sims}, were calibrated to achieve systematically lower cluster gas fractions, either $2\sigma$ (dashed grey) or $4\sigma$ (dashed black) below the observational data. These shifts primarily alter the AGN parameters of the model, although the SN parameters are varied as well \citep{kugel23}. We see that the astrophysical parameters have a very similar effect as the dark matter lifetime up to about $k\approx\SIF{1\hub}{\per\Mpc}$ at $z=0$, but that the effect of the lifetime is much stronger on smaller scales. This finding applies not just to the dark matter lifetime, but also to shifts in other cosmological parameters, as we will show below. Moreover, the effects of cosmological and astrophysical parameters also diverge over time. The right panel of Fig.~\ref{fig:dcdm_agn_variations} shows that the cosmological coupling is much weaker at $z=1$, relative to the impact of the astrophysical parameters. Hence, the degeneracy may be broken by considering the scale and time dependence of the baryonic suppression. While encouraging, baryonic processes are complex and by no means exhaustively described by the $f_\text{gas}$ variations in the FLAMINGO suite. Hence, further work is needed to disentangle the effects of cosmology and astrophysics.

\section{Halo model approaches}\label{section:approximations}

Baryonic effects are often included in cosmic shear analyses using approximate prescriptions that do not fully capture the dependence on cosmological parameters, with the result that feedback parameters encode both cosmological and astrophysical information. To illustrate the consequences, we will consider in Section~\ref{sec:hmcode} the commonly-used and simplest version of the halo model approach of \citet{mead15}, as implemented in \textsc{hmcode} and used in the KiDS-1000 analysis of \citet{asgari22}. In Section~\ref{sec:hmcode2020}, we will turn to a more sophisticated version that has been released since \citep{mead21}. This version provides a clear improvement, but still does not fully capture the cosmological coupling discussed in this paper.

\subsection{Comparison with \textsc{hmcode}}\label{sec:hmcode}

\begin{figure}
     \normalsize
     \centering
     \subfloat{
         \includegraphics{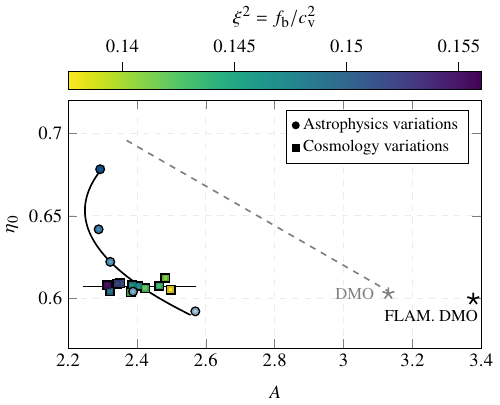}
     }\vspace{-0.5em}
     \caption{Best-fitting values of the two parameters, $A$ and $\eta_0$, used in the \textsc{hmcode} model to describe the effect of baryons on the matter power spectrum, obtained by fitting the model to the suppression of the power spectrum in the FLAMINGO simulations. The circles correspond to simulations that vary astrophysical parameters at fixed cosmology, while the squares correspond to simulations that vary cosmological parameters with fixed subgrid parameters. The grey star labelled DMO marks the parameters in the absence of baryonic effects and the grey dashed line is a fit, $\eta_0 = 0.98 - 0.12A$, from \citet{joudaki18} to feedback variations from the OWLS suite \citep{schaye10}. The black star marks the best-fiting parameters for the fiducial gravity-only FLAMINGO simulation.}
     \label{fig:hmcode_projection}
\end{figure}

Let us briefly review the basic \textsc{hmcode} approach, while referring to \citet{asgari23} for a more comprehensive overview. In the halo model, the matter power spectrum is written as the sum
\begin{align}
    P_\mathrm{m}(k) = P_{2\mathrm{h}}(k) + P_{1\mathrm{h}}(k),
\end{align}

\noindent
where the `2-halo' term, $P_{2\mathrm{h}}(k)$, describes the variance in the mass-weighted density of halos and the `1-halo' term, $P_{1\mathrm{h}}(k)$, describes the variance in the mass density of matter within halos. The model has a number of free parameters affecting both terms, but baryonic effects only impact the 1-halo term in this model. The 1-halo term is given by an integral over halo masses,
\begin{align}
    P_{1\mathrm{h}}(k) = \frac{1}{\bar{\rho}^2}\int_0^\infty M^2\tilde{\rho}^2(k,M) F(M)\mathrm{d}M, \label{eq:halo_model_1h}
\end{align}
\noindent
where $\bar{\rho}$ is the mean density of matter, $\tilde{\rho}(k,M)$ is the normalized Fourier transform of the density profile of a halo of mass $M$, and $F(M)$ is the halo mass function. In \textsc{hmcode}, the halo density profile is given by a modified NFW form \citep{navarro96}. The impact of baryons on $\tilde{\rho}$ is captured by two free parameters: $A$ and $\eta_0$. The first parameter, $A$, is a normalization of the relation between the concentration, $c$, of a halo and its mass, $M$, at redshift $z$:
\begin{align}
    c(M,z) = A\frac{1+z_\mathrm{f}}{1+z}, \label{eq:hmcode_concentrations}
\end{align}

\noindent
where $z_\mathrm{f}$ is the formation redshift of the halo. The second parameter, $\eta_0$, describes a mass-dependent modification of the density profile,
\begin{align}
    \tilde{\rho}(k,M) \to \tilde{\rho}(\nu^{\eta} k,M), \label{eq:profile_mod}
\end{align}

\noindent
where $\eta=\eta_0-0.3\sigma_8(z)$ and $\nu=\delta_\mathrm{c}/\sigma(M)$ is the peak height in terms of the linear-theory collapse threshold, $\delta_\mathrm{c}\approx1.686$, and the standard deviation of the linear matter density field, $\sigma(M)$, in spheres of radius $(3M/4\pi\bar{\rho})^{1/3}$. Fitting the model to power spectra from dark matter only simulations, \citet{mead15} determined best-fitting values of $(A,\eta_0)=(3.13, 0.603)$ in the absence of baryonic effects. This point is marked by a grey star labelled DMO in Fig.~\ref{fig:hmcode_projection}. Doing the same for power spectra measured from hydrodynamical simulations from the OWLS suite \citep{schaye10}, they obtained a range of parameters for different feedback scenarios. To eliminate one degree of freedom, \citet{joudaki18} and \citet{asgari22} exploited the correlation between $\eta_0$ and $A$ in the best-fitting OWLS parameters and used a linear fit, $\eta_0 = 0.98 - 0.12A$, in their cosmic shear analyses. This relation is shown as the grey dashed line in Fig.~\ref{fig:hmcode_projection}.

One of the advantages of the halo model approach is that it automatically captures some of the ways in which baryonic feedback could be coupled with cosmological parameters, namely through changes in the halo mass function, $F(M)$, or halo formation epochs, $z_\mathrm{f}$. However, the halo model does not capture the cosmology dependence of baryonic processes inside the galaxy. We have shown in preceding sections that this coupling causes properties like the masses of supermassive black holes (Eq.~\ref{eq:mbh_scaling} and Fig.~\ref{fig:bh_fb_scaling}) or gas escape fractions (Eq.~\ref{eq:ftilde} and Fig.~\ref{fig:fbar_clusters}) to depend on cosmological parameters.

\begin{figure}
     \normalsize
     \centering
     \subfloat{
         \includegraphics{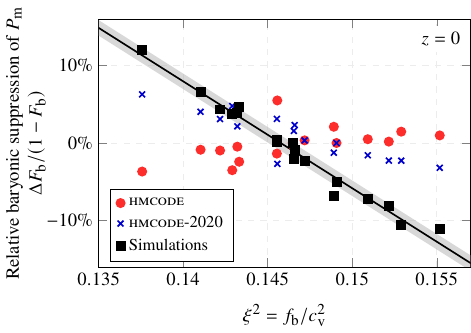}
     }\vspace{-0.5em}
	 \caption{Non-factorizable corrections to the power spectrum at $z=0$, relative to the fiducial L1\_m9 model, measured from the FLAMINGO simulations (black squares), and computed with \textsc{hmcode} (red circles) and \textsc{hmcode-2020} (blue crosses). The FLAMINGO points are the same as in the right-hand panel of Fig.~\ref{fig:pk_vs_c}.}
     \label{fig:suppression_vs_hmcode}
\end{figure}

To demonstrate the impact, we fit the model to the baryonic suppression, $F_\mathrm{b}(k)$, in our fiducial simulation (L1\_m9) with $A$ and $\eta_0$ as free parameters. We fit to the data at $z=0$ and use the implementation of \textsc{hmcode} in \textsc{class}. We then fix $A$ and $\eta_0$ at these best-fitting values and use \textsc{hmcode} to compute the non-factorizable corrections, $\Delta F_\mathrm{b}/(1-F_\mathrm{b})$, relative to the fiducial model, when the cosmological parameters are changed to those of the $\Lambda$CDM simulations discussed in Section~\ref{sec:additional_sims} and those listed in Table~\ref{tab:sims}. The average corrections, on scales $\SIF{0.1\hub}{\per\Mpc}\leq k \leq \SIF{10\hub}{\per\Mpc}$, are shown in red in Fig.~\ref{fig:suppression_vs_hmcode}, along with the corrections computed from the FLAMINGO simulations in black. We see that the cosmological coupling is much smaller in \textsc{hmcode} than in the simulations and, unlike for the simulations, there is no clear trend with respect to $\xi^2$. The agreement with the simulations is better for the more recent implementation, \textsc{hmcode-2020}, as will be discussed below.

Since \textsc{hmcode} does not fully capture the non-factorizable corrections seen in the simulations, the baryonic parameters $A$ and $\eta_0$ must depend on cosmological parameters. To demonstrate this, we fit the model to $F_\mathrm{b}(k)$ and obtain the best-fitting values of $A$ and $\eta_0$ for all of our simulations. The results are shown in Fig.~\ref{fig:hmcode_projection}, where the coloured squares correspond to the cosmology variations discussed in Section~\ref{sec:additional_sims} and the blue circles correspond to the astrophysics variations L1\_m9\_$f_\text{gas}\pm n\sigma$ listed in Table~\ref{tab:sims}. These astrophysics variations have subgrid physics parameters calibrated to cluster gas fractions that are $n\sigma$ higher or lower than the observed data \citep{kugel23}, but all assume the fiducial cosmology. Interestingly, the two types of variations appear to have different effects on $\eta_0$ and $A$, with $\eta_0$ being essentially independent of cosmology but quite sensitive to shifts in astrophysical parameters. This can be understood in terms of a difference in scales: $\eta_0$ affects the suppression on larger scales, while $A$ is more important on smaller scales. The finding that the cosmological coupling is more important on small scales agrees with what was seen in Section~\ref{section:s8_degeneracies} for the decaying dark matter models. We note that \textsc{hmcode} treats $A$ and $\eta_0$ as free parameters that are chosen to match the power spectrum, rather than the actual halo concentrations and density profiles in simulations. Hence, the dependence of $A$ on cosmological parameters cannot be interpreted directly in terms of the halo concentration. Instead, the figure demonstrates that $A$ depends on the combination $\xi^2=f_\mathrm{b}/c_\mathrm{v}^2$.

Next, let us compare the astrophysics variations from the FLAMINGO suite with the grey dashed line, which corresponds to the linear fit based on simulations from the OWLS suite \citep{schaye10}. In both cases, $\eta_0$ increases and $A$ generally decreases with the strength of baryonic feedback. However, the detailed behaviour is very different. For the same value of $A$, we find a smaller suppression and hence a smaller value of $\eta_0$ for the FLAMINGO simulations. This shows that a 1-parameter model is not sufficiently flexible to describe the baryonic suppression in general. Moreover, some cosmological information is lost when marginalizing over feedback parameters without fully modelling the cosmological coupling.

To mitigate these limitations, one could adopt more flexible modelling approaches that include the dependence on cosmology. For instance, one could incorporate Eq.~\eqref{eq:phys_model} into existing approximate prescriptions. As a first attempt, let us demonstrate how this could be done for the basic version of \textsc{hmcode}. Let $\bar{A}$ and $\bar{\eta}_0$ be the parameters that describe feedback for some fixed cosmology. Combining a quadratic fit, $\bar{A}(\bar{\eta}_0)$, to the astrophysics variations with fiducial cosmology and a linear fit, $A(\xi^2)$, to the cosmology variations with fiducial feedback parameters shown in Fig.~\ref{fig:hmcode_projection}, we obtain\footnote{We remind the reader that the baryon fraction, $f_\mathrm{b}=\Omega_\mathrm{b}/(\Omega_\mathrm{b}+\Omega_\mathrm{c})$, should be defined relative to the cold matter species, thus excluding neutrinos.}
\begin{align}
    \begin{split}
	A &= 36.69\left(1 - 2.738\bar{\eta}_0 + 2.096\bar{\eta}^2_0 - 0.2988\xi^2\right),
    \end{split}\label{eq:hmcode_fit}
\end{align}

\noindent
for $\bar{\eta}_0\in[0.59, 0.68]$ and $\xi^2 = f_\mathrm{b}/c^2_\mathrm{v}\cong 0.65 f_\mathrm{b}/(\Omega_\text{m}\sigma_8)^{1/4}$. This prescription covers a wide range of astrophysical and cosmological scenarios with a single feedback parameter that is independent of cosmology at first order: $\bar{\eta}_0$. The range $\bar{\eta}_0\in[0.59, 0.68]$ could be taken as a prior motivated by FLAMINGO. For a typical Planck cosmology, the model then excludes the DMO case with $(A,\eta_0)=(3.13, 0.603)$.

While the \textsc{hmcode} parameters are poorly constrained by current weak lensing observations, at least when assuming $\Lambda$CDM \citep[e.g.][]{joudaki18,asgari22}, this situation will change with the arrival of Euclid, LSST, and WFIRST, or when considering extensions like decaying dark matter. We caution that approximations like Eq.~\eqref{eq:hmcode_fit} are not adequate for future surveys \citep[e.g.][]{taylor18,lacasa19}. A more accurate approach would be to rely on suites of simulations that vary both astrophysical and cosmological parameters, like the CAMELS suite \citep{villaescusa21}, but in significantly larger volumes to model clustering accurately over the range of scales probed by weak lensing surveys.

\subsection{Comparison with \textsc{hmcode-2020}}\label{sec:hmcode2020}

Next, we will compare with an updated version of \textsc{hmcode} called \textsc{hmcode-2020} \citep{mead21}, which brings numerous improvement over the model discussed in the previous subsection. Let us focus on the implementation of baryonic feedback. As in the simpler model, \textsc{hmcode-2020} allows baryonic effects to alter halo concentrations by modifying the amplitude, $A$, of the mass-concentration relation \eqref{eq:hmcode_concentrations}. Baryonic effects also change the halo density profile, but rather than rescaling the profile as in Eq.~\eqref{eq:profile_mod}, it is modified via
\begin{align}
	\tilde{\rho}(k,M) \to \left(\frac{\Omega_\mathrm{c}}{\Omega_\mathrm{m}} + f_\mathrm{g}(M)\right)\tilde{\rho}(k,M) + f_*, \label{eq:hmcode2020_profile}
\end{align}
\noindent
where $f_\mathrm{g}(M)$ is the gas mass fraction and $f_*$ the stellar mass fraction. In this expression, the first term accounts for the expulsion of gas by rescaling the overall amplitude of the profile. Secondly, a constant shot noise term is added to model the stellar population. The mass-dependent gas fraction is defined as
\begin{align}
   f_\mathrm{g}(M) = \left(\frac{\Omega_\mathrm{b}}{\Omega_\mathrm{m}}-f*\right)f_\text{ret}(M),
\end{align}
\noindent
where $f_\text{ret}(M)$ is the fraction of gas that is retained, given by
\begin{align}
  f_\text{ret}(M) = 1 - f_\text{esc}(M) = \frac{(M/M_\mathrm{b})^2}{1+(M/M_\mathrm{b})^2}. \label{eq:hmcode_fesc}
\end{align}
\noindent
The parameter $M_\mathrm{b}$ is a transition mass, defined such that halos with $M\gg M_\mathrm{b}$ retain all their gas. Baryonic feedback is thus controlled by three parameters: $A, M_\mathrm{b}$, and $f_*$. In \citet{mead21}, three further parameters are used to model the time dependence of $A, M_\mathrm{b}$, and $f_*$.

To compare \textsc{hmcode-2020} with our results, we again fit the model to the baryonic suppression, $F_\mathrm{b}(k)$, at $z=0$ measured from our fiducial L1\_m9 simulations. We subsequently take the best-fitting values of $A, M_\mathrm{b},f_*$ as fixed, while varying the cosmological parameters, and use \textsc{hmcode-2020} to determine the non-factorizable corrections. The results are shown in Fig.~\ref{fig:suppression_vs_hmcode} as blue crosses. In contrast to the \textsc{hmcode} model discussed in the previous subsection, there is now a clear trend with higher baryon fractions, $f_\mathrm{b}$, and lower concentrations, $c_\mathrm{v}$, leading to stronger baryonic feedback. This is in qualitative agreement with our results. However, as can be seen from the figure, \textsc{hmcode-2020} predicts a weaker dependence on cosmology than what is seen in the FLAMINGO simulations, particularly for models with large values of $\xi^2$ and strong baryonic feedback. This is consistent with the comparison in \citet{mead21} of their model with the BAHAMAS simulations \citep{mccarthy17}.

We can understand this by noting that the escape fraction, $f_\text{esc}(M)$, given by Eq.~\eqref{eq:hmcode_fesc}, does not depend on cosmological parameters. This is in contrast to the discussion in Section~\ref{sec:towards}, where we argued that the escape fraction also depends on cosmology. In other words, while Eq.~\eqref{eq:hmcode2020_profile} accounts for the fact that a greater fraction of the mass is susceptible to feedback when $\Omega_\mathrm{b}/\Omega_\mathrm{m}$ is larger, it does not incorporate the finding that this feedback itself is also stronger. Similarly, the escape fraction should also depend on cosmological parameters through the halo concentration. In summary, \textsc{hmcode-2020} provides a clear improvement over \textsc{hmcode}, but does not fully capture the cosmological coupling discussed in this paper.

\section{Conclusion}\label{sec:discussion}

Upcoming weak lensing surveys will measure the clustering of matter with unprecedented precision and provide tight constraints on cosmological models. Since a large fraction of the information resides on non-linear scales, where baryonic processes are important, one crucial question for the interpretation of these observations is the coupling between the cosmological parameters of interest and astrophysical feedback processes. Previous work has often neglected this coupling, assuming the effects of cosmology and feedback to be independent. Using the FLAMINGO suite of hydrodynamical simulations \citep{kugel23,schaye23}, we determined that the cosmological coupling of baryonic feedback gives rise to non-factorizable corrections to the matter power spectrum on scales $\SIF{1\hub}{\per\Mpc}<k<\SIF{10\hub}{\per\Mpc}$, which cannot be ignored at the level of $1\%$ accuracy needed for Stage-IV surveys \citep[e.g.][]{taylor18,lacasa19}.

Our model of the non-factorizable corrections is built on a careful analysis of the ways in which baryonic feedback could depend on cosmology, using secondary halo properties, such as the halo concentration and formation epoch, as proxies for different mediating mechanisms. We determined that the cosmological effects on the baryon fractions and binding energies of halos are most important, with greater baryon fractions leading to stronger feedback, while greater binding energies lead to weaker feedback. By contrast, the role of the halo environment is negligible. We then constructed a simple physical model of AGN feedback and showed that the non-factorizable corrections to the power spectrum can be accurately predicted from a single parameter combination, $f_\mathrm{b}/c^2_\mathrm{v}\sim f_\mathrm{b}/(\Omega_\text{m}\sigma_8)^{1/4}$, where $f_\mathrm{b}$ is the universal baryon fraction and $c^2_\mathrm{v}$ a velocity-based definition of halo concentration, as described by equations \eqref{eq:phys_model2} and \eqref{eq:phys_model_cosmo}. Interestingly, our model predicts that feedback is stronger for models with lower $\sigma_8$ and $\Omega_\text{m}$, as suggested by the $S_8$ tension between high- and low-redshift probes of matter clustering \citep{planck18,asgari22,amon22,secco22,abbott23}. Given that baryonic and novel non-baryonic suppression mechanisms have both been considered as possible solutions to the $S_8$ tension \citep[e.g.][]{amon22b,preston23,mccarthy23}, the question arises whether the two could work together to produce a greater overall suppression.

By running hydrodynamical simulations for two models with suppressed structure formation, involving massive neutrinos or decaying dark matter, we demonstrate that the combined effect of baryonic and non-baryonic suppression mechanisms is indeed greater than the sum of its parts. For massive neutrinos, we find that a combination of a summed neutrino mass of $\sum m_\nu\in[0.24,0.48]\,\si{\eV}$ and standard baryonic feedback can reconcile CMB and weak lensing observations. However, this explanation is at odds with constraints on the expansion history inferred from CMB and BAO \citep{planck18} and other probes \citep[e.g.][]{palanque20,brieden22}. For decaying dark matter, we find a strong dependence of feedback on the dark matter lifetime, leading to sizeable non-factorizable corrections. For the most extreme model with a decay rate of $\Gamma=\SI{24}{\km/\s/\Mpc}$ ($\tau=\SI{41}{\Gyr}$), the baryonic suppression is roughly twice as strong as for the case without decaying dark matter. Such short dark matter lifetimes are similarly disfavoured by geometric constraints \citep[e.g][]{audren14,aubourg15}. Hence, both models can reconcile CMB and weak lensing measurements and relieve the $S_8$ tension, but only at the expense of distorting the expansion history. Nevertheless, it seems likely that other suppression mechanisms also enhance baryonic feedback. In particular, we expect that velocity-dependent dark matter annihilation \citep{choquette16} would give rise to larger corrections and a greater boost of baryonic feedback for the same overall loss of dark matter and change in expansion history, given that the loss would be concentrated in the centres of groups and clusters. This possibility will be explored in future works.

Finally, we considered the extent to which cosmological and astrophysical parameters are degenerate within the FLAMINGO model. We determined that the cosmological coupling of feedback becomes important on scales $k>\SIF{1\hub}{\per\Mpc}$ and for $z<1$, suggesting that the degeneracy with key parameters of the FLAMINGO galaxy formation model may be broken by considering the dependence of the power spectrum on scale and time. This would require a model for the redshift evolution of the non-factorizable corrections, which we leave as a topic for future work. We also looked at popular feedback prescriptions that do not fully account for the non-factorizable corrections \citep{mead15,mead21} and demonstrated that, as a result, their parameters encode both cosmological and astrophysical information. Hence, some cosmological information is lost by marginalizing over such parameters, which is the current practice for cosmic shear analyses. This is particularly problematic for non-standard parameters with greater effects on feedback, such as the neutrino mass and dark matter lifetime. We conclude that optimal use of forthcoming observations for cosmological inference requires baryonic feedback prescriptions that are more flexible and incorporate the dependence of feedback on cosmology or the use of large simulations that vary both astrophysical and cosmological parameters.

\section*{Acknowledgements}

We gratefully acknowledge detailed and insightful comments by the anonymous referee, which significantly improved the manuscript. WE, CSF, AJ, and BL acknowledge STFC Consolidated Grant ST/X001075/1 and support from the European Research Council through ERC Advanced Investigator grant, DMIDAS [GA 786910] to CSF. JB and RK acknowledge support from research programme Athena 184.034.002 from the Dutch Research Council (NWO). This project has received funding from the ERC under the European Union’s Horizon 2020 research and innovation programme (grant agreement No 769130). This work used the DiRAC@Durham facility managed by the Institute for Computational Cosmology on behalf of the STFC DiRAC HPC Facility (www.dirac.ac.uk). The equipment was funded by BEIS capital funding via STFC capital grants ST/K00042X/1, ST/P002293/1 and ST/R002371/1, Durham University and STFC operations grant ST/R000832/1. DiRAC is part of the National e-Infrastructure.

\section*{Data availability}

The FLAMINGO data will be released publicly in the future. Those interested in working with the simulations now are encouraged to contact the corresponding author.




\bibliographystyle{mnras}
\bibliography{main}



\appendix

\section{Dependence on simulation volume}\label{sec:convergence}

\begin{figure}
    \normalsize
    \centering
	\subfloat{
		\includegraphics{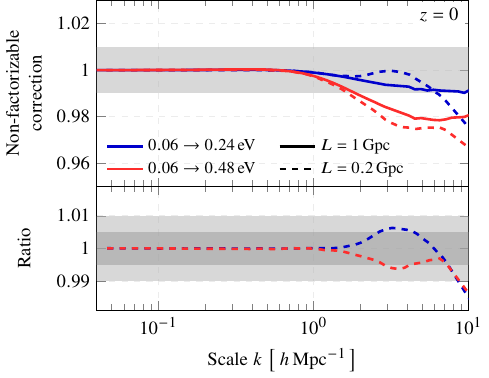}
	}\vspace{-0.5em}
    \caption{Dependence of the non-factorizable corrections on the simulation volume. We show the corrections at $z=0$ when changing the sum of neutrino masses from $\sum m_\nu=\SI{0.06}{\eV}$ to $\SI{0.24}{\eV}$ or $\SI{0.48}{\eV}$. Solid lines correspond to $(\SI{1}{\Gpc})^3$ simulations, while dashed lines correspond to $(\SI{0.2}{\Gpc})^3$ simulations. The ratio is mostly within $\pm0.5\%$ (dark grey) on the scales of interest.}
    \label{fig:separable_convergence}
\end{figure}

In this work, we primarily use two simulation volumes: most of the large simulations listed in Table~\ref{tab:sims} have side length $L=\SI{1}{\Gpc}$, while the smaller simulations described in Section~\ref{sec:additional_sims} have side length $L=\SI{200}{\Mpc}$. In Fig.~\ref{fig:separable_convergence}, we show the non-factorizable correction to the matter power spectrum at $z=0$ when changing the neutrino mass from $\sum m_\nu=\SI{0.06}{\eV}$ to $\SI{0.24}{\eV}$ or $\SI{0.48}{\eV}$, for both simulation volumes. We see that the two volumes agree to within about $0.5\%$ on the scales of interest. Since we are interested in effects exceeding $1\%$, we choose to use both volumes in our analysis, but expect to see a somewhat larger scatter for the power spectrum predictions from the smaller simulations.

\section{Additional fits}\label{sec:power_law_fits}

\begin{figure}
    \normalsize
    \centering
    \subfloat{
        \includegraphics{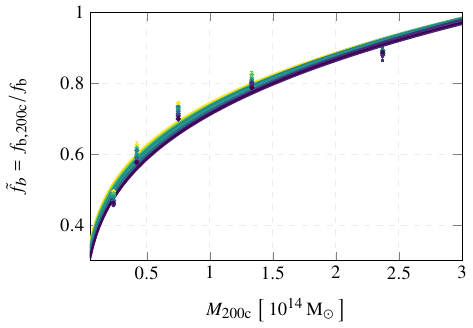}
    }
    \caption{Fits of the renormalized baryon fractions, $\tilde{f}_\mathrm{b}=f_{\mathrm{b},200\mathrm{c}}/f_\mathrm{b}$, where $f_{\mathrm{b},200\mathrm{c}}$ is the baryon fraction within $R_{200\mathrm{c}}$ for matched halos and $f_\mathrm{b}$ is the universal baryon fraction, as a function of halo mass, $M_{200\mathrm{c}}$, for 11 small $\Lambda$CDM simulations. The error bars are difficult to see except at the high-mass end. The points and curves are colour coded according to the average baryonic suppression of the power spectrum, as in Fig.~\ref{fig:separable_accuracy}.}
    \label{fig:fbar_clusters_mass_fits}
\end{figure}

In Section~\ref{sec:towards}, we introduced a two-parameter power law \eqref{eq:fb_power_law_fit} to describe the renormalized baryon fractions in halos, $\tilde{f}_\mathrm{b}=f_{\mathrm{b},200\mathrm{c}}/f_\mathrm{b}$, as a function of halo mass, $M_{200\mathrm{c}}$. In Fig.~\ref{fig:fbar_clusters_mass_fits}, we show these fits for the $\Lambda$CDM simulations from Section~\ref{sec:additional_sims}. The points correspond to the median of $\tilde{f}_\mathrm{b}$ in each mass bin with bootstrapped error bars that are difficult to see at the scale of the plot, except at the high-mass end. The simulations have the same colours as in Fig.~\ref{fig:separable_accuracy}, indicating the average baryonic suppression of the power spectrum. As expected, the unambiguous trend is that the power spectrum is more strongly suppressed for models with lower renormalized baryon fractions. The renormalized baryon fractions, in turn, are lower for simulations with greater universal baryon fractions, $f_\mathrm{b}$, and lower halo concentrations, $c_\mathrm{v}$, as shown in Fig.~\ref{fig:fbar_clusters}.

\begin{figure}
    \normalsize
    \centering
    \subfloat{
        \includegraphics{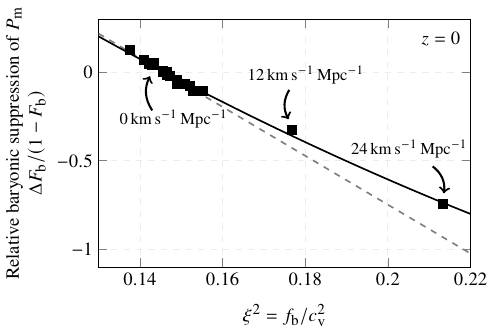}
    }
    \caption{Non-factorizable corrections to the power spectrum at $z=0$ as a function of $\xi^2$ for the FLAMINGO simulations with decaying dark matter (points labelled with decay rates of $\SI{12}{\km\per\s\per\Mpc}$ and $\SI{24}{\km\per\s\per\Mpc}$) and without decays (labelled with $\SI{0}{\km\per\s\per\Mpc}$). The fit of Eq.~\eqref{eq:phys_model2} in the right-hand panel of Fig.~\ref{fig:pk_vs_c} is reproduced here as a grey dashed line. A new two-parameter fit to all simulations, including those with decaying dark matter and given by Eq.~\eqref{eq:phys_model3}, is shown as a black line.} 
    \label{fig:suppression_with_dcdm}
\end{figure}

Next, we provide an extension of the right-hand panel of Fig.~\ref{fig:pk_vs_c}, showing the non-factorizable corrections to the power spectrum as a function of $\xi^2=f_\mathrm{b}/c_\mathrm{v}^2$. In Fig.~\ref{fig:suppression_with_dcdm}, we reproduce all the data points from Fig.~\ref{fig:pk_vs_c} and add two further points for the FLAMINGO simulations with decaying dark matter. The halos in these simulations are much more strongly affected by baryonic feedback and their concentrations and baryon fractions are far outside the parameter ranges considered in Fig.~\ref{fig:pk_vs_c}. We caution that there is an additional choice in the definition of $\xi^2$ for the decaying dark matter simulations, since the universal baryon fraction, $f_\mathrm{b}$, now depends on time. In this plot, we have chosen to evaluate $f_\mathrm{b}$ at $z=0$, given that we do the same for $c_\mathrm{v}$ and $P(k)$.

The one-parameter fit to the $\Lambda$CDM (and massive neutrino) simulations from Eq.~\eqref{eq:phys_model2} is shown as a grey dashed line in Fig.~\ref{fig:suppression_with_dcdm}. When extrapolated, this model does not fit the decaying dark matter simulations well, but this is not surprising given that the models are far outside the original parameter range. We can describe all simulations, including the ones with decaying dark matter, by adding just one further parameter:
\begin{align}
\frac{\Delta F_\mathrm{b}}{1-F_\mathrm{b}} &= -\alpha_1\Delta(\xi^2) + \alpha_2\left[\Delta(\xi^2)\right]^2 \label{eq:phys_model3},
\end{align}
\noindent
with $\alpha_1=18.1\pm3.5$ and $\alpha_2=19.9\pm10.1$. This relation is shown as a black curve in Fig.~\ref{fig:suppression_with_dcdm}.

\begin{figure}
    \normalsize
    \centering
    \subfloat{
        \includegraphics{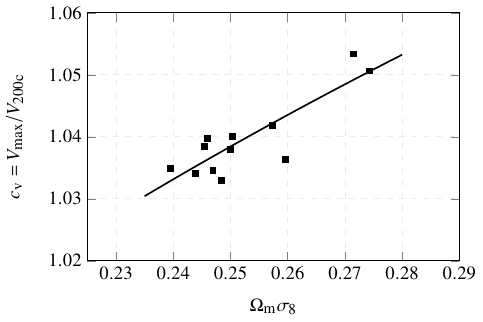}
    }
    \caption{Fit of the median halo concentration, $c_\mathrm{v}$, in terms of the cosmological parameter combination $\Omega_\mathrm{m}\sigma_8$, as given by Eq.~\eqref{eq:concentration_fit}.}
    \label{fig:c_tilde_fit}
\end{figure}

Finally, we show a fit of the median halo concentration, $c_\mathrm{v}$, for halos with masses $M_{200\mathrm{c}}$ in the range $[10^{13.75}\mathrm{M}_\odot, 10^{14.25}\mathrm{M}_\odot]$, as a function of the cosmological parameter combination $(\Omega_\mathrm{m}\sigma_8)^{1/8}$. This parametrization was introduced in Eq.~\eqref{eq:concentration_fit} in Section~\ref{section:s8_ranges}. The data for the $\Lambda$CDM simulations from Section~\ref{sec:additional_sims} and the resulting fit are shown in Fig.~\ref{fig:c_tilde_fit}.

\label{lastpage}
\end{document}